\documentclass[fleqn,usenatbib]{mnras}

\usepackage{times}
\usepackage[T1]{fontenc}
\usepackage{graphicx}
\usepackage{amsmath}
\usepackage{amssymb}
\usepackage{flushend}

\newcommand{\Msun}{\ensuremath{\,{\rm M}_\odot}}                  
\newcommand{\Rsun}{\ensuremath{\,{\rm R}_\odot}}                  
\newcommand{\Msunnom}{\hbox{$\mathcal{M}^{\rm N}_\odot$}}
\newcommand{\Rsunnom}{\hbox{$\mathcal{R}^{\rm N}_\odot$}}
\newcommand{\Lsunnom}{\hbox{$\mathcal{L}^{\rm N}_\odot$}}
\newcommand{\Teff}{\ensuremath{T_{\rm eff}}}                      
\newcommand{\kms}{\,km\,s$^{-1}$}                                 
\newcommand{\as}{\ensuremath{^{\prime\prime}}}                    

\newcommand{\kepler}{\textit{Kepler}}

\newcommand{\reff}[1]{#1}
\newcommand{\refff}[1]{#1}

\title[High-mass pulsators in eclipsing binaries]
      {High-mass pulsators in eclipsing binaries observed using TESS}

\author[Southworth \& Bowman]
       {John Southworth\,$^{1}$, Dominic M.\ Bowman\,$^2$ \\
        $^1$\,Astrophysics Group, Keele University, Staffordshire, ST5 5BG, UK \\
        $^2$\,Institute of Astronomy, KU Leuven, Celestijnenlaan 200D, B-3001 Leuven, Belgium
        }

\date{Accepted 2022 March 28. Received 2022 March 11; in original form 2021 December 14.}
\pubyear{2022}

\begin{document} \label{firstpage} \pagerange{\pageref{firstpage}--\pageref{lastpage}} \maketitle 

\begin{abstract}
\reff{Pulsations and binarity are both common features of massive stars. The study of pulsating massive stars in eclipsing binary systems hold great potential} for constraining stellar structure and evolution theory. However, prior to the all-sky Transiting Exoplanet Survey Satellite (TESS) mission, few such systems had been discovered or studied in detail. We have inspected the TESS light curves of a large number of eclipsing binaries known to contain high-mass stars, and compiled a list of 18 objects which show intrinsic variability. The light curves were modelled both to determine the physical properties of the systems, and to remove the effects of binarity in order to leave residual light curves suitable for asteroseismic analysis. Precise mass and radius measurements were obtained for $\delta$\,Cir, CC\,Cas, SZ\,Cam V436\,Per and V539\,Ara. We searched the residual light curves for pulsation signatures and\reff{, within our sample of 18 objects,} we find \refff{six definite and eight possible cases of $\beta$\,Cephei pulsation,} seven cases of stochastic low-frequency (SLF) variability, and \refff{eight} instances of possible slowly pulsating B (SPB) star pulsation. The large number of pulsating eclipsing systems we have identified makes asteroseismology of high-mass stars in eclipsing binaries a feasible avenue to constrain the interior physics of a large sample of massive stars for the first time.
\end{abstract}



\section{Introduction}
\label{sec:intro}

High-mass stars are preferentially found in binary and multiple systems \citep{Sana+12sci,Sana+14apjs,Kobulnicky+14apjs} and a significant fraction have short orbital periods. They are also bright so are over-represented in magnitude-limited samples. As a result, a relatively large number of high-mass stars have been found to be members of eclipsing binaries (EBs). Of particular interest are high-mass stars in EBs with orbital periods long enough for them to have evolved as single stars. For these objects it is possible to measure their masses and radii directly and use these properties to compare with or to calibrate theoretical models of \reff{the} evolution of single stars \citep[e.g.][]{Andersen++90apj,Ribas++00mn,Torres++10aarv}. A prominent recent trend is the finding that the properties of high-mass EBs need stronger internal mixing processes than predicted by standard evolutionary models \citep{Tkachenko+20aa,Johnston21aa}. Subsequent evolution of high-mass short-period binaries leads to a wide variety of exotic objects that are important for many areas of stellar astrophysics, such as X-ray binaries, supernovae and gamma-ray bursts \citep{Podsiadlowski+02apj,Podsiadlowski+04apj,Belczynski+20aa,Chrimes+20mn}.

Many high-mass stars also show pulsations \citep{Bowman20faas}. Among early-type main-sequence stars of spectral type types O and B, there are three main classes of pulsating variable. The slowly pulsating B (SPB) stars are mid-to-late B stars that pulsate in high-radial order gravity (g) modes with periods of order days \citep{Waelkens91aa}. The $\beta$\,Cephei stars span spectral types from O9 to B3 on the main sequence but can reach B5 during the giant phase. They pulsate in low-radial order pressure (p) and g modes with periods of order a few hours \citep{StankovHandler05apjs}. The pulsation modes of SPB and $\beta$\,Cephei stars are driven by a heat-engine mechanism operating in the partial ionisation zones of iron and nickel at 200\,000~K \citep{DziembowskiPamyatnykh93mn,Dziembowski++93mn}. This opacity-based driving mechanism produces periodic standing waves (i.e.\ coherent pulsation modes), from which forward asteroseismic modelling can reveal a star's interior physics \citep{Aerts++10book,Aerts21rvmp}. See \citet{Bowman20faas} for a recent review of forward asteroseismic results of SPB and $\beta$\,Cephei stars. In particular, the combination of dynamical masses and radii from binary modelling with asteroseismic modelling of pulsations in eclipsing systems has shown great promise in being able to precisely constrain stellar structure and evolution theory (see e.g.\ \citealt{SchmidAerts16aa,Johnston+19mn}). But such studies are so far lacking among massive stars.

The third type of pulsator among early-type stars are those which exhibit stochastic low-frequency (SLF) variability \citep{Bowman+19natas,Bowman+19aa,Bowman+20aa}. Such stars have quasi-periodic and time-dependent variability spanning a broad range of periods from several days to of order minutes, which has been inferred to be caused by stochastically excited gravity waves driven by turbulent (core) convection \citep{Bowman+19aa}. These gravity waves are an efficient mixing mechanism inside massive stars \citep{RogersMcelwaine17apj}, and are seemingly ubiquitous in massive stars with precise enough photometry \citep{Bowman+19natas}. Gravity waves are excited at the interface of convective and radiative regions inside massive stars, which include the \reff{convective} core as predicted by two- and three-dimensional hydrodynamical simulations \citep{Rogers+13apj,Edelmann+19apj,Horst+20aa}. Another explanation for SLF variabilty in massive stars is caused by the dynamics of their turbulent envelopes, which also gives rise to temperature and velocity variations \citep{Cantiello+21apj,Schultz+22apjl}. A comparison of recent high-cadence TESS photometry and high-resolution spectroscopy allowed \citet{Bowman+20aa} to demonstrate that the SLF variability in massive stars probes the mass and age of a star, and offers further support to the conclusion that gravity waves are responsible for macroturbulence in massive stars \citep{Aerts+09aa,Simondiaz+10apj,Grassitelli+15apj,Grassitelli+16aa,Simondiaz+17aa}. Therefore the continued study of SLF variability, especially in binary systems that have so far been neglected, is an exciting prospect for constraining the interior mixing of massive stars as so far their study has been mainly based on single stars. To do so, a well-characterised sample of massive binary systems is needed, which is one of the motivations of this work.

\reff{Here} we present the discovery and preliminary analysis of a set of pulsating high-mass stars in EBs. A detailed study of these systems in most cases will require extensive spectroscopy and sophisticated analysis, which we leave to future work. We have also determined the physical properties of targets for which suitable data were available.


\section{Target selection}
\label{sec:targ}

Only a few EBs containing high-mass pulsators were known until recently. The main reason for this is that the amount of data needed to securely detect pulsations photometrically is much higher compared to their single-star counterparts. Furthermore, disentangling the photometric signals of binarity, rotation and pulsations of massive stars require ultra-high photometric precision, which is difficult to achieve from the ground. SPB and $\beta$\,Cephei stars are observationally difficult to detect because of their low pulsation amplitudes (certainly compared to eclipse depths); SPB stars also have pulsation periods longer than a single night so are difficult to study using ground-based observations. We note that some high-mass pulsators in EBs have been identified spectroscopically (e.g.\ V2107\,Cyg; \citealt{Bakis+14aj}) through line-profile variations (LPVs).

The \kepler\ space mission brought a huge improvement in the quality and quantity of time-series photometry for pulsating stars, but its small sky coverage meant few high-mass stars were observed and even fewer high-mass pulsators in EBs were identified. The prime example of this is the discovery of SLF variability in the B1.5\,III + B2\,V system V380\,Cyg \citep{Tkachenko+12mn,Tkachenko+14mn}. The appearance of TESS \citep{Ricker+15jatis} has changed this picture completely as it is in the process of obtaining time-series high-precision photometry of a large fraction of the sky. In contrast to \kepler, the TESS mission has observed a large number of high-mass EBs, of which many have an extensive observational history and in some cases a detailed characterisation of the physical properties of the component stars. See the recent study by \citet{IJspeert+21aa} for a census of EBs with g-mode pulsations predominantly focussed on intermediate-mass stars, \citet{Prsa+21apjs} for a catalogue of 4500 EBs detected using TESS, and \citet{Me21univ} for a detailed review of the impact of space-based telescopes on binary star science.

We therefore searched the TESS database to find EBs containing high-mass pulsating stars. The starting list was taken to be a bibliography of objects gradually accumulated by the first author since the year 2001, and therefore includes only those EBs identified as such in a past publication. Each one was entered in the data visualisation portal at the Mikulski Archive for Space Telescopes (MAST) archive\footnote{\texttt{https://mast.stsci.edu/portal/Mashup/Clients/ Mast/Portal.html}} and any available short-cadence light curves were visually inspected for relevant features. This target selection was performed manually in summer 2020 -- many more of these objects will be detectable using automated methods and subsequent TESS data.

\reff{The full set of 26 objects identified in our search is given in Table\,\ref{tab:targ}. The TESS data of four of these (V453\,Cyg, CW\,Cep, VV\,Ori and V\,Pup) have already been the subject of a detailed analysis so are not studied further in the current work. A further four are currently undergoing analysis by ourselves or collaborators, so here we announce the discovery of pulsations but do not perform any analysis. The remaining 18 objects are discussed individually below. For completeness, we include all of the objects we identified as high-mass EBs containing pulsating stars in Table\,\ref{tab:targ}.}

\citet{Labadie+20aj} recently presented the discovery of five new $\beta$\,Cephei EBs using photometry from the KELT project \citep{Pepper+07pasp}. The TESS light curves of these objects confirm the nature in three cases (V447\,Cep, HD\,339003 and HD\,344880). A fourth object, HD\,227977, shows $\beta$\,Cephei pulsations but not eclipses in the TESS light curve so is unlikely to be a $\beta$\,Cephei EB. The fifth star, HD\,254346, has no TESS light curve but was considered to be a probable blend rather than a $\beta$\,Cephei EB \citep{Labadie+20aj}. We have not considered these objects further in the current work, as \citet{Labadie+20aj} state they are already following them up.

\begin{table*}
\caption{\label{tab:targ} Basic information for the targets considered in this work. For simplicity they are given in alphabetical order, beginning with
numbers then Greek letters then Latin letters. The spectral types come from a variety of sources, with ones from refereed journal articles preferred.}
\begin{tabular}{@{}lcclllll@{}} \hline
Target        & HD     & $V$       & Spectral type                  & Orbital    & Pulsation type                 & TESS          & Analysis of TESS data         \\
              & number & magnitude &                                & period (d) &                                & sector(s)     &                               \\
\hline
16 Lac        & 216916 &   5.59    & B2\,IV                         & 12.097     & $\beta$\,Cephei                & 16            & This work                     \\
$\delta$ Cir  & 135240 &   5.04    & O7\,III-V + O9.5\,V + B0.5\,V  & 3.902      & SLF                            & 12            & This work                     \\
$\eta$ Ori    &  35411 &   3.34    & B1\,V + B2:                    & 7.988      & $\beta$\,Cephei                & 6, 32         & This work                     \\
$\lambda$ Sco & 158926 &   1.62    & B1.5\,IV + PMS + B2\,V         & 5.953      & $\beta$\,Cephei                & 12, 39        & This work                     \\
$\mu$ Eri     &  30211 &   4.01    & B5\,IV                         & 7.381      & SLF + SPB?                     & 5, 32         & This work                     \\
AN Dor        &  31407 &   7.69    & B2/3\,V                        & 2.033      & \refff{$\beta$\,Cephei/SPB}    & 2--6, 29--32  & This work                     \\
AR Cas        & 221253 &   4.89    & B3\,V                          & 6.067      & SPB                            & 17, 24        & Southworth et al.\ (in prep.) \\
AS Cam        &  35311 &   8.60    & A0\,V                          & 3.431      & SPB                            & 19, 52        & Southworth et al.\ (in prep.) \\
CC Cas        &  19820 &   7.10    & O8.5\,III + B0.5\,V            & 3.366      & SLF                            & 18, 19        & This work                     \\
CW Cep        & 218066 &   7.64    & B1.5\,Vn                       & 2.729      & $\beta$\,Cephei                & 17, 18, 24    & \citet{LeeHong21aj}           \\
EM Car        &  97484 &   8.51    & O7.5\,V((f)) + O7.5\,V((f))    & 3.415      & SLF                            & 10, 11, 37    & Torres et al.\ (in prep.)     \\
EO Aur        &  34333 &   7.76    & B0\,V + B3\,V                  & 4.065      & $\beta$\,Cephei                & 19            & This work                     \\
FZ CMa        &  42942 &   8.09    & B2\,V + B2\,V                  & 1.273      & \refff{$\beta$\,Cephei/SPB}    & 7, 33         & This work                     \\
HD 217919     & 217919 &   8.27    & B0.5\,III                      & 16.206     & \refff{$\beta$\,Cephei/SPB}    & 17, 18, 24    & This work                     \\
HQ CMa        &  57593 &   5.99    & B3\,V                          & unknown    & $\beta$\,Cephei/SPB            & 7, 34, 35     & This work                     \\
LS CMa        &  52670 &   5.64    & B2-3\,III-IV                   & 70.23      & $\beta$\,Cephei? SPB?          & 6, 7, 33, 34  & This work                     \\
QX Car        &  86118 &   6.64    & B2\,V                          & 4.478      & $\beta$\,Cephei?               & 9, 10, 36, 37 & Torres et al.\ (in prep.)     \\
SZ Cam        &  25638 &   6.93    & O9\,IV + B0.5\,V               & 2.698      & SLF                            & 19            & This work                     \\
V Pup         &  65818 &   4.47    & B1\,Vp + B2:                   & 1.454      & $\beta$\,Cephei                & 7, 9, 34, 35  & \citet{Budding+21mn}          \\
V379 Cep      & 197770 &   6.31    & B2\,IV-III                     & 99.764     & SLF                            & 15--17        & This work                     \\
V436 Per      &  11241 &   5.53    & B1.5\,V                        & 25.936     & SLF                            & 18            & This work                     \\
V446 Cep      & 210478 &   7.73    & B1\,V + B9\,V                  & 3.808      & $\beta$\,Cephei + TEO?         & 16, 17, 24    & This work                     \\
V453 Cyg      & 227696 &   8.40    & B0.4\,IV + B0.7\,IV            & 3.890      & $\beta$\,Cephei                & 14, 15, 41    & \citet{Me+20mn}               \\
V539 Ara      & 161783 &   5.68    & B3\,V + B4\,V                  & 3.169      & SLF + SPB?                     & 13, 39        & This work                     \\
V2107 Cyg     & 191473 &   8.63    & B1\,III                        & 4.284      & \refff{$\beta$\,Cephei/SPB}    & 14, 15, 41    & This work                     \\
VV Ori        &  36695 &   5.34    & B1\,V + B7\,V                  & 1.495      & $\beta$\,Cephei + SPB          & 6, 32         & \citet{Me++21mn}              \\
\hline
\end{tabular}
\end{table*}

\section{Observations}
\label{sec:obs}

TESS was launched by NASA on 2018/04/08 into an eccentric orbit around the Earth resonant with the orbit of the Moon \citep{Ricker+15jatis}. It is currently performing a photometric survey of 85\% of the celestial sphere with the aim of identifying extrasolar planets through the transit method \citep{Ricker+15jatis}. The observational equipment comprises four cameras with 10.5\,cm apertures, each with four CCDs, that together image a contiguous 24$^\circ$$\times$96$^\circ$ strip of sky. The pixel size on the sky is 21$^{\prime\prime}$$\times$21$^{\prime\prime}$, so contamination occurs for many stars. TESS observes through a wide-band filter that covers approximately 600\,nm to 1000\,nm.

Each strip of sky is observed by TESS for two orbits (27.4\,d), with a break near the midpoint for downlink of data to Earth. Each set of observations is called a sector, and some sky areas are observed in multiple sectors. A total of 200\,000 stars were pre-selected for high-cadence observations (summed into a 120\,s sampling rate on board the satellite). Full-frame images are also captured at a cadence of 1800\,s (again summed on board), and subsequently clipped into an effective integration time of 1425\,s by a cosmic-ray rejection algorithm.

The data are processed and released as light curves by the TESS Science Processing Operations Center (SPOC) for stars observed at high cadence \citep{Jenkins+16spie}. The standard data product is simple aperture photometry (SAP). An additional data product, pre-search data conditioning (PDC) light curves, is available which has been processed to make it more suitable for detecting shallow transits of extrasolar planets in front of their host stars. The PDC data are not intended for stars which show large variability amplitudes, such as many EBs, and in our experience often contain strong systematics introduced by the conditioning process. We therefore used \reff{exclusively the SAP data} in the current work. These were downloaded from the MAST, extracted from the fits files, and converted into magnitude units. We rejected any datapoints having a nonzero flag for the QUALITY parameter, with the exception of a few cases noted below. We also ignored the errorbars provided as these are usually far too small and the data are relatively homogeneous, preferring instead to determine the quality of our modelling from the scatter of the data around the best fits. Light curves for each of our targets are shown in Figs.\ \ref{fig:time:1},  \ref{fig:time:2} and \ref{fig:time:3}.


\section{Analysis methods}
\label{sec:anal}

\subsection{Analysis of eclipses}
\label{sec:anal:ecl}

The first analysis step was to attempt to model the effects of binarity in the TESS light curve. This served dual aims: to measure the radii and orbital inclination of the component stars, and to subtract the signatures of binarity from the light curve prior to investigation of the pulsations. For this step we used version 42 of the {\sc jktebop}\footnote{\texttt{http://www.astro.keele.ac.uk/jkt/codes/jktebop.html}} code \citep{Me++04mn2,Me13aa}. This code treats the stars as spheres for the calculation of eclipse shapes and as ellipsoids for the simulation of the ellipsoidal effect. Such an approximation is suitable for systems where the stars are not very tidally deformed, and in the case of well-detached systems {\sc jktebop} has been shown to agree very well with other codes \citep{Maxted+20mn}. For close binaries where the components are significantly aspherical, {\sc jktebop} still provides a good fit to the data but the parameters of the fit become unreliable. In some cases we have continued to use {\sc jktebop} in preference to more sophisticated codes because we find that it is orders of magnitude faster both for the user and the computer; we give the approximate parameters of the system for reference but leave detailed modelling to the future.

Throughout this paper we refer to the primary star (the one obscured at the deeper eclipse) as star~A. The secondary star is star~B and any tertiary star (if present) is called star~C.

In {\sc jktebop} the radii of the stars are parameterised in terms of the fractional radii, defined by $r_{\rm A} = \frac{R_{\rm A}}{a}$ and $r_{\rm B} = \frac{R_{\rm B}}{a}$ where $R_{\rm A}$ and $R_{\rm B}$ are the true radii of the stars, and $a$ is the semimajor axis of the relative orbit. The fitted parameters are the sum ($r_{\rm A}+r_{\rm B}$) and ratio ($k = \frac{r_{\rm B}}{r_{\rm A}}$) of the radii, as these are more closely related to the eclipse shapes. We also fitted for the orbital inclination ($i$), the ratio of the central surface brightnesses of the stars ($J$), the orbital period ($P$) and the time of midpoint of a primary eclipse ($T_0$). Limb darkening was included using the quadratic law \citep{Kopal50} with theoretical coefficients for the TESS passband from \citet{Claret17aa}; we typically fitted for the linear coefficients of the two stars ($u_{\rm A}$ and $u_{\rm B}$) and fixed the quadratic coefficients ($v_{\rm A}$ and $v_{\rm B}$) because of the strong correlations between $u$ and $v$ \citep{Me++07aa}. In cases where the orbit was eccentric we additionally fitted the quantities $e\cos\omega$ and $e\sin\omega$, where $e$ is the orbital eccentricity and $\omega$ is the argument of periastron. Some systems required the inclusion of third light ($\ell_3$) as a fitted parameter. Finally, we fitted for the coefficients of low-order polynomials applied to the out-of-eclipse brightness of the systems as a function of time in order to remove any slow variations in brightness due to either astrophysical or instrumental effects.

\subsubsection{Wilson-Devinney code}

Some of the systems studied in this work have stars sufficiently close to each other to be significantly tidally distorted. In these cases a spherical-star approximation (as used in {\sc jktebop}) is not suitable and full Roche geometry is needed. Where the prospect of extracting useful information was large, we fitted the light curve with the Wilson-Devinney (WD) code \citep{WilsonDevinney71apj,Wilson79apj} in its 2004 version ({\sc wd2004}), driven by the {\sc jktwd} wrapper \citep{Me+11mn}. Prior to this, we converted each light curve to orbital phase and then binned it into 400 datapoints in order to save computing time. These data were fitted in mode 0 or 2, assuming (pseudo)synchronous rotation, logarithmic limb darkening using coefficients from \citet{Vanhamme93aj}, gravity darkening exponents of 1.0 as suitable for hot stars \citep{Claret98aas} and albedos of 1.0. Mass ratios and effective temperature (\Teff) values were taken from previous work and the Cousins $R$ band was adopted as a reasonable approximation to the TESS passband for hot stars.

The fitted parameters in each case were the potentials of both stars, the orbital inclination, a phase shift, the light contributions of the two stars, the amount of third light in the system, and where necessary $e$ and $\omega$. Third light is expressed in terms of the fractional contribution to the total light of the system at a phase of 0.25, so is not directly comparable to the light contributions from the two stars individually \citep{WilsonVanhamme04}.

Error analysis with the WD code is non-trivial because the formal errors from the covariance matrix are usually too small, especially for data of space-based quality \citep{Me20obs}. We therefore perturbed the fit by changing the rotation rates, albedo and gravity darkening exponents by $\pm$0.1, using a different limb darkening law, changing the mass ratio by its uncertainty, and trying different numerical resolutions (see \citealt{Pavlovski++18mn} and \citealt{Me20obs}). The uncertainties were obtained from the variations between the different models and the final adopted parameter values and in all cases are larger than the formal errors calculated by {\sc wd2004}.

\begin{table}
\caption{\label{tab:vsini} \refff{Published projected rotational velocity measurements, $V\sin i$,
for the target stars. In the case of EO\,Aur the FWHM of the line profiles was measured from
fig.\,2 in \citet{Popper78apj} in order to set some limit on the $V\sin i$ of this system.}}
\begin{tabular}{lccl} \hline
Target        & $V_1\sin i$ & $V_2\sin i$  & Reference \\
Target        & (\kms)      & (\kms)       &          \\
\hline
16 Lac        & approx.\ 40 &              & \citet{Aerts+01aa}         \\
$\delta$ Cir  & $150 \pm 9$ & $141 \pm 20$ & \citet{Penny+01apj}        \\ 
$\eta$ Ori    &      20     &      130     & \citet{Demey+96aa}         \\
$\lambda$ Sco &     140     &              & \citet{Uytterhoeven+04aa}  \\ 
$\mu$ Eri     & $136 \pm 6$ &              & \citet{Jerzykiewicz+13mn}  \\
CC Cas        & $336 \pm 5$ & $202 \pm 13$ & \citet{Hill+94aa}          \\
EO Aur        & $\ga$300   & $\ga$300    & this work, \citet{Popper78apj} \\
FZ CMa        & $216 \pm 10$& $216 \pm 10$ & \citet{Moffat+83aa}        \\
HQ CMa        &     130     &              & \citet{Wolff++82apj}       \\
LS CMa        &      17     &              & \citet{Abt++02apj}         \\
SZ Cam        & $144 \pm 3$ & $117 \pm 4$  & \citet{Tamajo+12aa}        \\ 
V379 Cep      &      55     &       15     & \citet{Gordon+98aj}        \\
V436 Per      &     115     &      140     & \citet{Janik+03aa}         \\
V446 Cep      & $120 \pm 3$ &  $44 \pm 9$  & \citet{Cakirli+14}         \\
V539 Ara      &  $75 \pm 8$ &  $48 \pm 5$  & \citet{Andersen83aa}       \\
V2107 Cyg     &  $84 \pm 4$ &              & \citet{Bakis+14aj}         \\
\hline
\end{tabular}
\end{table}

\subsection{Physical properties}
\label{sec:anal:phys}

Some objects have previously been studied spectroscopically, so their full physical properties can be calculated using the parameters measured from the light curve and published spectroscopic orbits. To do this we used the {\sc jktabsdim} code  \citep{Me++05aa}, which implements standard equations \citep[e.g.][]{Hilditch01book} and propagates errorbars using a perturbation analysis. We used the IAU nominal solar properties and physical constants \citep{Prsa+16aj} for compatibility with other work. We calculated the masses of the stars ($M_{\rm A}$ and $M_{\rm B}$), their radii ($R_{\rm A}$ and $R_{\rm B}$), surface gravities ($\log g_{\rm A}$ and $\log g_{\rm B}$), light ratio in the TESS passband ($\ell_{\rm B}/\ell_{\rm A}$) and the semimajor axis ($a$).

Where possible we adopted published values for the \Teff\ values of the stars in order to calculate their luminosities ($L_{\rm A}$ and $L_{\rm B}$). Distances to the systems were not considered as this is beyond the scope of the current work. \refff{At the request of the referee we have compiled published values of the projected rotational velocities of the systems for which this is available in Table\,\ref{tab:vsini}. The physical properties of the systems we determine} are given in Tables \ref{tab:lc:1}, \ref{tab:lc:2} and \ref{tab:lc:wd}.

\subsection{Pulsations}
\label{sec:anal:puls}

For the majority of systems, we used the residual light curve after subtracting the binary model to calculate an amplitude spectrum using a discrete Fourier Transform \citep{Kurtz85mn}. Similarly to \citet{Me+20mn,Me++21mn}, we identified significant \reff{and resolved (following $1/\Delta(T)$, where $\Delta(T)$ is the length of the TESS data set) pulsation frequencies and calculated their optimised parameters by using a non-linear least-squares cosinusoid fit} to the residual light curve. We used a signal-to-noise ratio (S/N) $\geqslant$ 5 as our significance criterion following previous studies of {\it Kepler} and TESS data that advocate using values larger than the canonical S/N $\geqslant$ 4 limit of \citet{Breger+93aa} (see e.g. \citealt{Burssens+20aa,BaranKoen21aca,BowmanMichielsen21aa}). However, some stars had large pulsation amplitudes relative to their eclipse depths. This made binary modelling difficult without first identifying the dominant pulsation mode frequencies. Therefore for such stars we manually clipped the eclipses out of the light curve before performing frequency analysis. These systems were: 16\,Lac, $\lambda$\,Sco and $\mu$\,Eri.

It is common that in the case of an imperfect binary model, residual power is found at integer multiples of the orbital frequency (i.e.\ harmonics) in the amplitude spectrum of a residual light curve. If a significant frequency is extracted at the location of an orbital harmonic we do not consider it an independent pulsation frequency. This is because our goal is to identify stars with coherent heat-driven pulsations for future asteroseismic modelling. However, it is certainly plausible that at least some of the significant frequencies that coincide with orbital harmonics are tidally-excited oscillation (TEO) modes if the binary system is sufficiently eccentric (see e.g.\ \citealt{Welsh+11apjs, Thompson+12apj}). However, differentiating TEO modes from residual amplitude from an imperfect binary model is beyond the scope of this work, as this requires more extensive spectroscopic monitoring and modelling. We are currently gathering spectroscopy of the most promising systems but leave the analysis for future work.

For each \refff{frequency} analysis, we produced a summary figure in the supplementary material, in which the top panel shows the residual light curve (i.e. after subtraction of the binary model) and the bottom panel shows the amplitude spectrum of the residual light curve with labelled significant pulsation mode frequencies. In the systems that had their eclipses removed prior to their frequency analysis, the significant frequencies are labelled in purple and are shown in Fig.\,A1. The summaries of other stars are shown in Figs.\ A2 and A3, and have their significant independent frequencies labelled in green. For each star, we also provide the pulsation mode frequencies, amplitudes and phases with uncertainties from the non-linear least-squares fit in Table\,A1. In Table\,\ref{tab:targ} we also provide a pulsator classification for each target based on visual inspection\reff{, which is based on the significant frequencies detected and the spectral type and/or mass of the star. However, in the case of pulsating binaries there is ambiguity of which star or stars in the system are pulsating.}


\section{Discussion of individual objects}

\begin{figure*} \includegraphics[width=\textwidth]{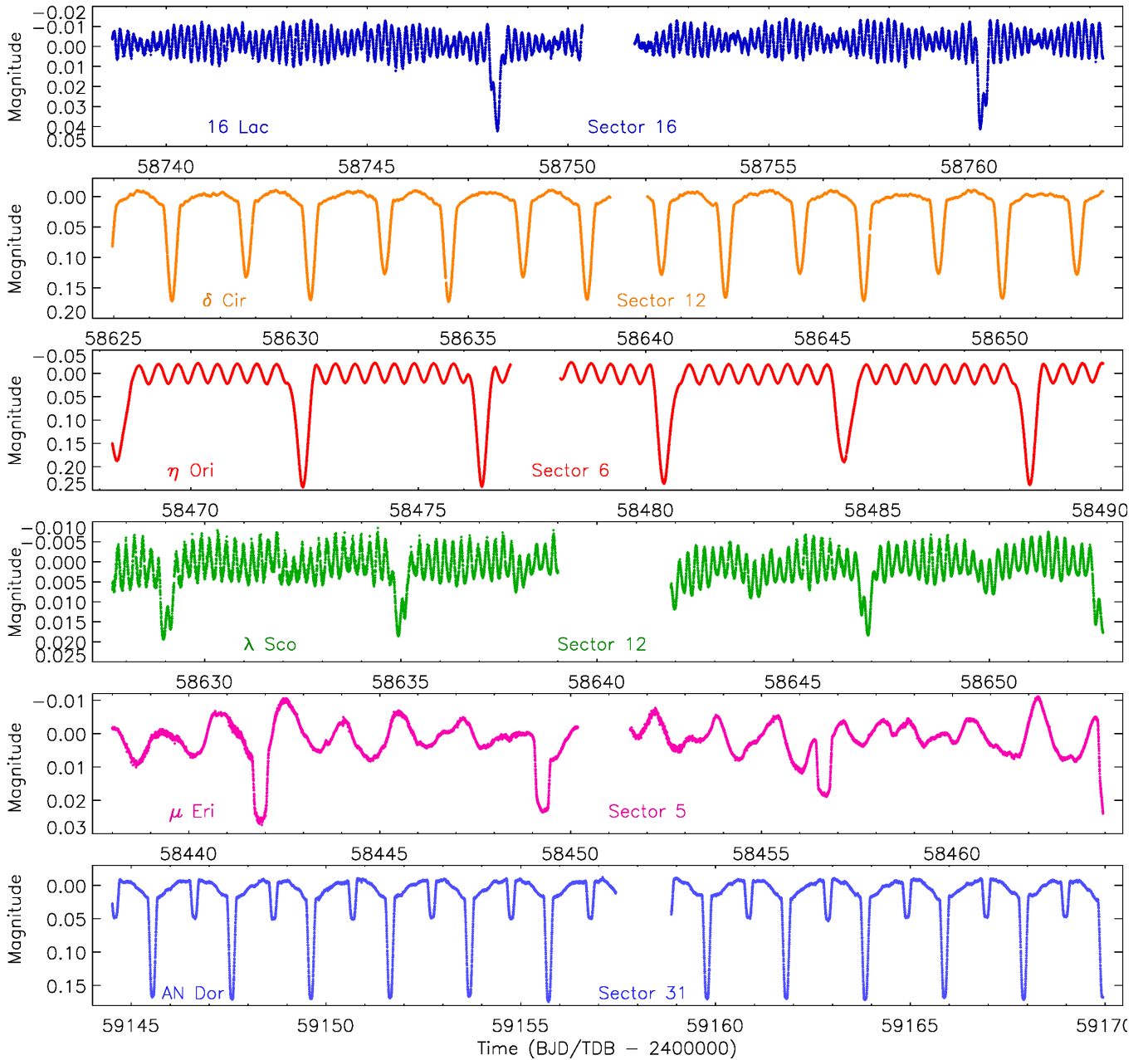}
\caption{\label{fig:time:1} TESS light curves of the first six objects
analysed in this work. In each case one sector is plotted. The object
names and the sectors are labelled on the diagram.} \end{figure*}

\begin{figure*} \includegraphics[width=\textwidth]{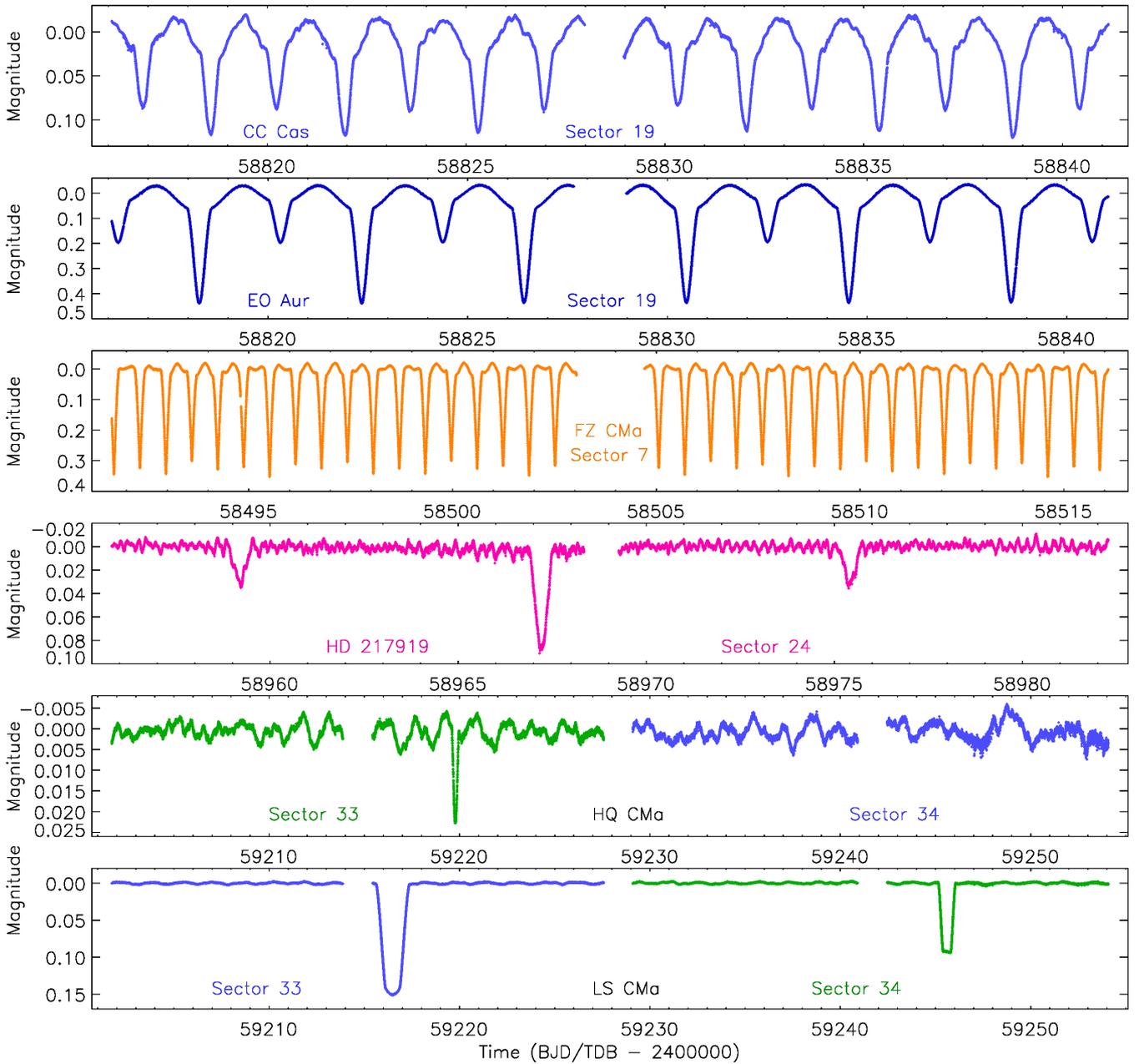}
\caption{\label{fig:time:2} TESS light curves of the middle six objects analysed in this
work. In four cases one sector is plotted, and in two cases two sectors are plotted.
The object names and the sectors are labelled on the diagram.} \end{figure*}

\begin{figure*} \includegraphics[width=\textwidth]{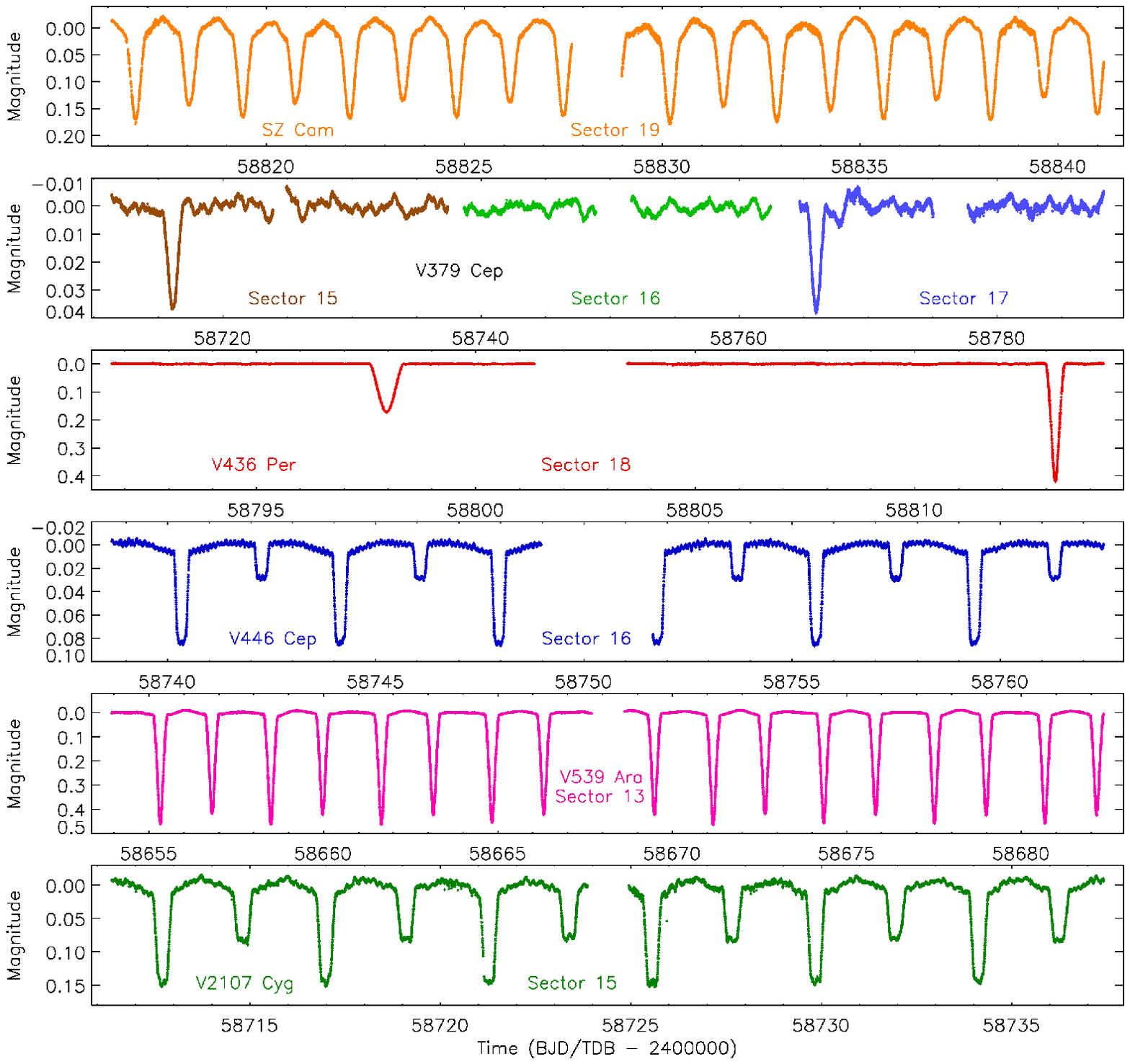}
\caption{\label{fig:time:3} TESS light curves of the last six objects
analysed in this work. In most cases one sector is plotted. The object
names and the sectors are labelled on the diagram.} \end{figure*}

\begin{table*} \centering
\caption{\label{tab:lc:1} Physical properties of those systems for which a detailed analysis with {\sc jktebop} was possible.
All times are given as BJD(TDB) $-$ 2400000.}
\setlength{\tabcolsep}{3pt}
\begin{tabular}{@{}lcccccc@{}}
\hline
                                              & 16 Lac                      & $\eta$ Ori                & LS CMa                  & V436 Per                    & V446 Cep                    & V539 Ara                    \\
\hline
\multicolumn{5}{@{}l}{\it Fitted parameters:} \\
$r_{\rm A}+r_{\rm B}$                         & $0.193 \pm 0.003$           & $0.2413 \pm 0.0042$       & $0.05779 \pm 0.00019$   & $0.08015 \pm 0.00028$       & $0.31624 \pm 0.00067$       & $0.3958 \pm 0.0013$         \\
$k$                                           & $0.52 \pm 0.02$             & $0.7395 \pm 0.0022$       & $0.36559 \pm 0.00070$   & $1.097 \pm 0.022$           & $0.25676 \pm 0.00042$       & $0.786 \pm 0.009$           \\
$i$ ($^\circ$)                                & $80.2 \pm 0.2$              & $87.62 \pm 0.42$          & $89.461 \pm 0.019$      & $87.951 \pm 0.025$          & $85.58 \pm 0.11$            & $85.16 \pm 0.13$            \\
$J$                                           & $0.18 \pm 0.05$             & $0.9044 \pm 0.0042$       & $0.6747 \pm 0.0056$     & $1.041 \pm 0.022$           & $0.4832 \pm 0.0040$         & $0.9601 \pm 0.0035$         \\
$\ell_3$                                      & 0.0 fixed                   & $0.468 \pm 0.016$         & 0.0 fixed               & $-0.003 \pm 0.011$          & 0.0 fixed                   & $0.0328 \pm 0.0028$         \\
$u_{\rm A}$                                   & 0.07 fixed                  & 0.00 fixed                & $0.402 \pm 0.081$       & $0.046 \pm 0.021$           & $0.0910 \pm 0.015$          & $0.146 \pm 0.023$           \\
$u_{\rm B}$                                   & 0.10 fixed                  & 0.00 fixed                & 0.10 fixed              & $= u_{\rm A}$               & 0.24 fixed                  & $= u_{\rm A}$               \\
$v_{\rm A}$                                   & 0.22 fixed                  & 0.35 fixed                & $-0.34 \pm 0.14$        & 0.22 fixed                  & 0.08 fixed                  & 0.21 fixed                  \\
$v_{\rm B}$                                   & 0.22 fixed                  & 0.35 fixed                & 0.22 fixed              & 0.22 fixed                  & 0.24 fixed                  & 0.21 fixed                  \\
$e\cos\omega$                                 & 0.017 fixed $^{(a)}$        & $-$0.00913$\pm$0.00010    & $-$0.12541$\pm$0.00013  & $-$0.12838$\pm$0.00011      & 0.00766$\pm$0.00011         & $-$0.04600$\pm$0.00097      \\
$e\sin\omega$                                 & 0.032 fixed $^{(a)}$        & 0.0027$\pm$0.0033         & $-$0.3281$\pm$0.0021    & 0.3614$\pm$0.0020           & $-$0.0119$\pm$0.0019        & $-$0.0311$\pm$0.0019        \\
$P$ (d)                                       & 12.095$\pm$0.002            & 7.98763$\pm$0.00032       & 70.02358$\pm$0.00034    & 25.935953$\pm$0.000034      & 3.808385$\pm$0.000004       & 3.169090$\pm$0.000021       \\
$T_0$                                         & 58748.220$\pm$0.002         & 58480.4275$\pm$0.0003     & 59216.4877$\pm$0.0015   & 58813.2011$\pm$0.0001       & 58767.0005$\pm$0.0001       & 58671.1670$\pm$0.0003       \\
$K_{\rm A}$ (\kms)                            & $23.818 \pm 0.033$ $^{(a)}$ & $145.5 \pm 0.03$ $^{(c)}$ &                         & $97.4 \pm 0.2$ $^{(d)}$     & $26600 \pm 1000$ $^{(e)}$   & $150.4 \pm 0.8$ $^{(f)}$    \\
$K_{\rm B}$ (\kms)                            &                             & $150 \pm 3$ $^{(c)}$      &                         & $91.2 \pm 0.2$ $^{(d)}$     & $22200 \pm 1000$            & $176.6 \pm 0.8$ $^{(f)}$    \\
\multicolumn{5}{@{}l}{\it Derived parameters:}\\
$r_{\rm A}$                                   & $0.1272 \pm 0.0005$         & $0.13871 \pm 0.00066$     & $0.04232 \pm 0.00015$   & $0.03822 \pm 0.00051$       & $0.25163 \pm 0.00050$       & $0.22171 \pm 0.00055$       \\
$r_{\rm B}$                                   & $0.0676 \pm 0.0030$         & $0.1026 \pm 0.0019$       & $0.01547 \pm 0.00005$   & $0.04191 \pm 0.00029$       & $0.06461 \pm 0.00020$       & $0.1742 \pm 0.0016$         \\
$e$                                           & 0.0392 $^{(a)}$             & $0.0095 \pm 0.0010$       & $0.3512 \pm 0.0020$     & $0.3835 \pm 0.0018$         & $0.0141 \pm 0.0016$         & $0.0555 \pm 0.0011$         \\
$\omega$ ($^\circ$)                           & 63.7 $^{(a)}$               & $164 \pm 18$              & $249.08 \pm 0.14$       & $109.5 6\pm 0.11$           & $302.8 \pm 4.3$             & $214.2 \pm 1.1$             \\
Light ratio                                   & $0.05 \pm 0.02$             & $0.496 \pm 0.022$         & $0.09094 \pm 0.00025$   & $1.253 \pm 0.032$           & $0.03186 \pm 0.00026$       & $0.5912 \pm 0.0057$         \\
$M_{\rm A}$ (\Msunnom)                        &                             & $10.87 \pm 0.44$          &                         & $6.880 \pm 0.037$           &                             & $6.239 \pm 0.066$           \\
$M_{\rm B}$ (\Msunnom)                        &                             & $10.54 \pm 0.22$          &                         & $7.348 \pm 0.039$           &                             & $5.313 \pm 0.060$           \\
$R_{\rm A}$ (\Rsunnom)                        &                             & $6.477 \pm 0.073$         &                         & $3.415 \pm 0.046$           &                             & $4.551 \pm 0.019$           \\
$R_{\rm B}$ (\Rsunnom)                        &                             & $4.79 \pm 0.10$           &                         & $3.745 \pm 0.027$           &                             & $3.575 \pm 0.035$           \\
$\log g_{\rm A}$ (c.g.s.)                     & $3.95 \pm 0.05$ $^{(b)}$    & $3.851 \pm 0.010$         &                         & $4.209 \pm 0.012$           &                             & $3.9170 \pm 0.0029$         \\
$\log g_{\rm B}$ (c.g.s.)                     & $3.50 \pm 0.04$             & $4.100 \pm 0.016$         &                         & $4.157 \pm 0.006$           &                             & $4.0570 \pm 0.0084$         \\
\Teff$_{\rm ,A}$ (K)                          & $23200 \pm 200$ $^{(b)}$    & 26600 $^{(c)}$            &                         & $21500 \pm 1000$ $^{(d)}$   &                             & $18100 \pm 500$ $^{(g)}$    \\
\Teff$_{\rm ,B}$ (K)                          & $14700 \pm 1200$            & 25950                     &                         & $22000 \pm 1000$ $^{(d)}$   &                             & $17100 \pm 500$ $^{(g)}$    \\
$\log(L_{\rm A}/\Lsunnom)$                    &                             &                           &                         & $3.351 \pm 0.081$           &                             & $3.302 \pm 0.048$           \\
$\log(L_{\rm B}/\Lsunnom)$                    &                             &                           &                         & $3.471 \pm 0.079$           &                             & $2.993 \pm 0.052$           \\
\hline
\end{tabular}
\newline
References:
$^{(a)}$ \citet{Lehmann+01aa};
$^{(b)}$ \citet{NievaPrzybilla12aa};
$^{(c)}$ \citet{Demey+96aa};
$^{(d)}$ Taken from \citet{Janik+03aa}. The errorbars for $K_{\rm A}$ and $K_{\rm B}$ have been doubled. No errorbars were given for the \Teff\ values so errorbars of 1000~K have been assumed. \\
$^{(e)}$ \citet{Cakirli+14};
$^{(f)}$ \citet{Andersen83aa};
$^{(g)}$ \citet{Clausen96aa};
\end{table*}

\begin{table*} \centering
\caption{\label{tab:lc:2} Physical properties of those systems for which an approximate analysis with {\sc jktebop} was performed.
Quantities in brackets are uncertainties in the final digits of the preceding numbers. All times are given as BJD(TDB) $-$ 2400000.}
\setlength{\tabcolsep}{4pt}
\begin{tabular}{@{}lccccccc@{}}
\hline
                                              & $\lambda$ Sco             & $\mu$ Eri                 & AN Dor          & EO Aur          & HD 217919      & V379 Cep                & V2107 Cyg                 \\
\hline
\multicolumn{5}{@{}l}{\it Fitted parameters:} \\
$r_{\rm A}+r_{\rm B}$                         & 0.32                      & 0.40                      & 0.40            & 0.44            & 0.17           & 0.072                   & 0.38                      \\
$k$                                           & 0.17                      & 0.28                      & 0.35            & 0.62            & 0.41           & 0.96                    & 0.34                      \\
$i$ ($^\circ$)                                & 80.0                      & 68.4                      & 80.8            & 83.1            & 83.2           & 86.9                    & 86.1                      \\
$J$                                           & 0.26                      & 0.15 fixed                & 0.051           & 0.76            & 0.80           & 0.96                    & 0.59                      \\
$\ell_3$                                      & 0.54 fixed                & 0.0 fixed                 & 0.0 fixed       & 0.0 fixed       & 0.20           & 0.50 fixed              & 0.0 fixed                 \\
$u_{\rm A}$                                   & 0.05 fixed                & 0.10 fixed                & 0.33            & 0.05 fixed      & 0.05 fixed     & 0.07 fixed              & 0.05                      \\
$u_{\rm B}$                                   & 0.24 fixed                & 0.18 fixed                & 0.24 fixed      & 0.25 fixed      & 0.28 fixed     & 0.07 fixed              & 0.10 fixed                \\
$v_{\rm A}$                                   & 0.15 fixed                & 0.22 fixed                & 0.10 fixed      & 0.09 fixed      & 0.07 fixed     & 0.22 fixed              & 0.22 fixed                \\
$v_{\rm B}$                                   & 0.25 fixed                & 0.29 fixed                & 0.21 fixed      & 0.21 fixed      & 0.20 fixed     & 0.22 fixed              & 0.21 fixed                \\
$e\cos\omega$                                 & 0.033                     &                           & 0.043           & 0.0 fixed       & 0.009          & 0.0 fixed               & $-0.015$                  \\
$e\sin\omega$                                 & $-$0.032                  &                           & $-$0.033        & 0.0 fixed       & 0.124          & 0.0 fixed               & 0.030                     \\
$P$ (d)                                       & 5.94501                   & 7.3813 (3)                & 2.032671 (1)    & 4.065556 (3)    & 16.20566 (5)   & 99.7638 fixed           & 4.284446 (15)             \\
$T_0$ (BJD/TDB)                               & 58634.974                 & 58449.2324 (3)            & 59155.71301 (1) & 58830.47762 (5) & 58805.1534 (3) & 58716.1004 (8)          & 58712.70759 (6)           \\
$K_{\rm A}$ (\kms)                            & $39.3 \pm 0.4$ $^{(a)}$   & $24.24 \pm 0.53$ $^{(b)}$ &                 &                 &                & 43.4 $^{(c)}$           & 104 $^{(d)}$              \\
$K_{\rm B}$ (\kms)                            &                           &                           &                 &                 &                & 79.7 $^{(c)}$           & 187 $^{(d)}$              \\
\multicolumn{5}{@{}l}{\it Derived parameters:}\\
$r_{\rm A}$                                   & 0.28                      & 0.31                      & 0.30            & 0.27            & 0.12           & 0.037                   & 0.29                      \\
$r_{\rm B}$                                   & 0.046                     & 0.09                      & 0.10            & 0.17            & 0.05           & 0.035                   & 0.097                     \\
$e$                                           & 0.047                     & 0.344 fixed $^{(b)}$      & 0.054           & 0.0             & 0.124          & 0.0                     & 0.034                     \\
$\omega$ ($^\circ$)                           & 315                       & 160.5 fixed $^{(b)}$      & 322             &                 & 85.8           &                         & 117                       \\
Light ratio                                   & 0.0072                    & 0.01                      & 0.0067          & 0.29            & 0.13           & 0.87                    & 0.066                     \\
$M_{\rm A}$ (\Msunnom)                        &                           &                           &                 &                 &                & 12.5                    & 7.1                       \\
$M_{\rm B}$ (\Msunnom)                        &                           &                           &                 &                 &                & 6.8                     & 3.9                       \\
$R_{\rm A}$ (\Rsunnom)                        &                           &                           &                 &                 &                & 8.9                     & 7.1                       \\
$R_{\rm B}$ (\Rsunnom)                        &                           &                           &                 &                 &                & 8.5                     & 2.4                       \\
$\log g_{\rm A}$ (c.g.s.)                     &                           & $3.55 \pm 0.04$ $^{(b)}$  &                 &                 &                & 3.6                     & 3.6                       \\
$\log g_{\rm B}$ (c.g.s.)                     & 4.44                      &                           &                 &                 &                & 3.4                     & 4.3                       \\
\Teff$_{\rm ,A}$ (K)                          & $25000 \pm 1000$ $^{(a)}$ & $15590 \pm 120$ $^{(b)}$  &                 &                 &                & 22000 $^{(c)}$          & $22500 \pm 1500$ $^{(d)}$ \\
\Teff$_{\rm ,B}$ (K)                          &                           &                           &                 &                 &                & 20200 $^{(c)}$          & $15200 \pm 1600$ $^{(d)}$ \\
$\log(L_{\rm A}/\Lsunnom)$                    &                           &                           &                 &                 &                &                         & 4.1                       \\
$\log(L_{\rm B}/\Lsunnom)$                    &                           &                           &                 &                 &                &                         & 2.4                       \\
\hline
\end{tabular}
\newline
References:
$^{(a)}$ \citet{Uytterhoeven+04aa};
$^{(b)}$ \citet{Jerzykiewicz+13mn};
$^{(c)}$ \citet{Harmanec+07aa};
$^{(d)}$ \citet{Bakis+14aj}.
\end{table*}

\begin{table*} \centering
\caption{\label{tab:lc:wd} Physical properties of those systems for which an analysis with {\sc jktwd} was performed.
All times are given as BJD(TDB) $-$ 2400000. 
}
\setlength{\tabcolsep}{5pt}
\begin{tabular}{@{}lcccccc@{}}
\hline
                                                & $\delta$ Cir             & CC Cas                   & FZ CMa                & SZ Cam                      \\
\hline
\multicolumn{5}{@{}l}{\it Parameters from {\sc jktebop}:} \\
$P$ (d)                                         & $3.902493 \pm 0.000025$  & $3.366029 \pm 0.000024$  &$1.273071 \pm 0.000008$  & $2.69863 \pm 0.0003$      \\
$T_0$ (BJD/TDB)                                 & $58638.34812 \pm 0.00009$& $58811.83691 \pm 0.00013$&$58501.87752 \pm 0.00003$& $58827.49911 \pm 0.00009$ \\
\multicolumn{5}{@{}l}{\it WD control and fixed parameters:} \\
WD mode                                         & 2                        & 0                        & 2                     & 2                           \\
Treatment of reflection                         & 1                        & 1                        & 1                     & 1                           \\
Number of reflections                           & 1                        & 1                        & 1                     & 1                           \\
Limb darkening law                              & logarithmic              & logarithmic              & logarithmic           & logarithmic                 \\
Numerical grid size (normal)                    & 40                       & 40                       & 60                    & 60                          \\
Numerical grid size (coarse)                    & 30                       & 30                       & 50                    & 60                          \\
\Teff\ of star~A (K)                            & 37500                    & 34500 $^{(b)}$           & 22000 $^{(d)}$        & $30320 \pm 150$ $^{(e)}$    \\
\Teff\ of star~B (K)                            & n/a                      & 29000 $^{(b)}$           & n/a                   & $28015 \pm 130$ $^{(e)}$    \\
Mass ratio ($M_2/M_1$)                          & $0.570 \pm 0.008$        & $0.424 \pm 0.009$        & $0.991 \pm 0.073$     & $0.747 \pm 0.006$           \\
Rotation rate for star~A                        & 1.0                      & 1.0                      & 1.0                   & 1.0                         \\
Rotation rate for star~B                        & 1.0                      & 1.0                      & 1.0                   & 1.0                         \\
Albedo for star~A                               & 1.0                      & 1.0                      & 1.0                   & 1.0                         \\
Albedo for star~B                               & 1.0                      & 1.0                      & 1.0                   & 1.0                         \\
Gravity darkening exponent for star~A           & 1.0                      & 1.0                      & 1.0                   & 1.0                         \\
Gravity darkening exponent for star~B           & 1.0                      & 1.0                      & 1.0                   & 1.0                         \\
Logarithmic LD coefficient star~A               & 0.2353                   & 0.2465                   & 0.2070                & 0.2597                      \\
Logarithmic LD coefficient star~B               & 0.2097                   & 0.2124                   & 0.2032                & 0.2374                      \\
\multicolumn{5}{@{}l}{\it WD fitted parameters:} \\
Potential of star~A                             & $4.204 \pm 0.029$        & $3.284 \pm 0.038$        & $4.55 \pm 0.07$       & $3.43 \pm 0.22$             \\
Potential of star~B                             & $4.815 \pm 0.067$        & $3.181 \pm 0.057$        & $4.57 \pm 0.20$       & $4.22 \pm 0.48$             \\
Orbital inclination ($^\circ$)                  & $77.80 \pm 0.19$         & $65.44 \pm 0.21$         & $88.05 \pm 0.20$      & $83.0 \pm 1.9$              \\
Orbital eccentricity                            & $0.05868 \pm 0.0048$     & $0.0099 \pm 0.0013$      & 0.0                   & 0.0                         \\
Argument of periastron ($^\circ$)               & $353.6 \pm 4.3$          & $209 \pm 27$             & n/a                   & n/a                         \\
Light from star~A                               & $7.35 \pm 0.28$          & $9.67 \pm 0.36$          & $3.511 \pm 0.084$     & $4.75 \pm 0.59$             \\
Light from star~B                               & n/a                      & $2.48 \pm 0.14$          & n/a                   & n/a                         \\
\Teff\ of star~B (K)                            & $30400 \pm 400$          & n/a                      & $20241 \pm 125$       & $29910 \pm 430$             \\
Linear LD coefficient star~A                    & 0.3794 (fixed)           & $0.52 \pm 0.14$          & $0.000 \pm 0.086$     & $0.457 \pm 0.033$           \\
Linear LD coefficient star~B                    & 0.3561 (fixed)           & $0.42 \pm 0.24$          & $0.037 \pm 0.084$     & $0.445 \pm 0.072$           \\
Third light                                     & $0.277 \pm 0.023$        &                          & $0.4667 \pm 0.0061$   & $0.531 \pm 0.046$           \\
$K_{\rm A}$ (\kms)                              & $153.0 \pm 1.4$ $^{(a)}$ & $123.9 \pm 2.0$ $^{(c)}$ & $215 \pm 8$ $^{(d)}$  & $185.2 \pm 2.8$ $^{(e)}$    \\
$K_{\rm B}$ (\kms)                              & $268.2 \pm 2.8$ $^{(a)}$ & $292.4 \pm 4.6$ $^{(c)}$ & $217 \pm 11$ $^{(d)}$ & $247.0 \pm 3.6$ $^{(e)}$    \\
\multicolumn{5}{@{}l}{\it Derived parameters:}\\
$r_{\rm A}$                                     & $0.2812 \pm 0.0018$      & $0.3570 \pm 0.0043$      & $0.2857 \pm 0.0005$   & $0.3706 \pm 0.0093$         \\
$r_{\rm B}$                                     & $0.1618 \pm 0.0025$      & $0.2245 \pm 0.0055$      & $0.2822 \pm 0.0026$   & $0.2087 \pm 0.0092$         \\
$M_{\rm A}$ (\Msunnom)                          & $20.00 \pm 0.50$         & $23.49 \pm 0.92$         & $5.16 \pm 0.56$       & $13.22 \pm 0.47$            \\
$M_{\rm B}$ (\Msunnom)                          & $11.41 \pm 0.24$         & $9.95 \pm 0.34$          & $5.11 \pm 0.46$       & $9.91 \pm 0.35$             \\
$R_{\rm A}$ (\Rsunnom)                          & $9.256 \pm 0.091$        & $10.87 \pm 0.18$         & $3.001 \pm 0.094$     & $8.61 \pm 0.24$             \\
$R_{\rm B}$ (\Rsunnom)                          & $5.326 \pm 0.091$        & $6.84 \pm 0.18$          & $2.964 \pm 0.097$     & $4.85 \pm 0.22$             \\
$\log g_{\rm A}$ (c.g.s.)                       & $3.806 \pm 9,997$        & $3.736 \pm 0.013$        & $4.196 \pm 0.022$     & $3.689 \pm 0.023$           \\
$\log g_{\rm B}$ (c.g.s.)                       & $4.043 \pm 0.014$        & $3.766 \pm 0.022$        & $4.203 \pm 0.018$     & $4.063 \pm 0.039$           \\
$\log(L_{\rm A}/\Lsunnom)$                      & $5.184 \pm 0.070$        & $5.179 \pm 0.053$        & $3.28 \pm 0.16$       & $4.752 \pm 0.025$           \\
$\log(L_{\rm B}/\Lsunnom)$                      & $4.339 \pm 0.090$        & $4.474 \pm 0.064$        & $3.12 \pm 0.19$       & $4.116 \pm 0.040$           \\
\hline
\end{tabular}
\newline
References:
$^{(a)}$ \citet{Penny+01apj};
$^{(b)}$ \citet{Hill+94aa};
$^{(c)}$ \citet{Gorda13astbu};
$^{(d)}$ \citet{Moffat+83aa};
$^{(e)}$ \citet{Tamajo+12aa}.
\end{table*}

\begin{figure*} \includegraphics[width=\textwidth]{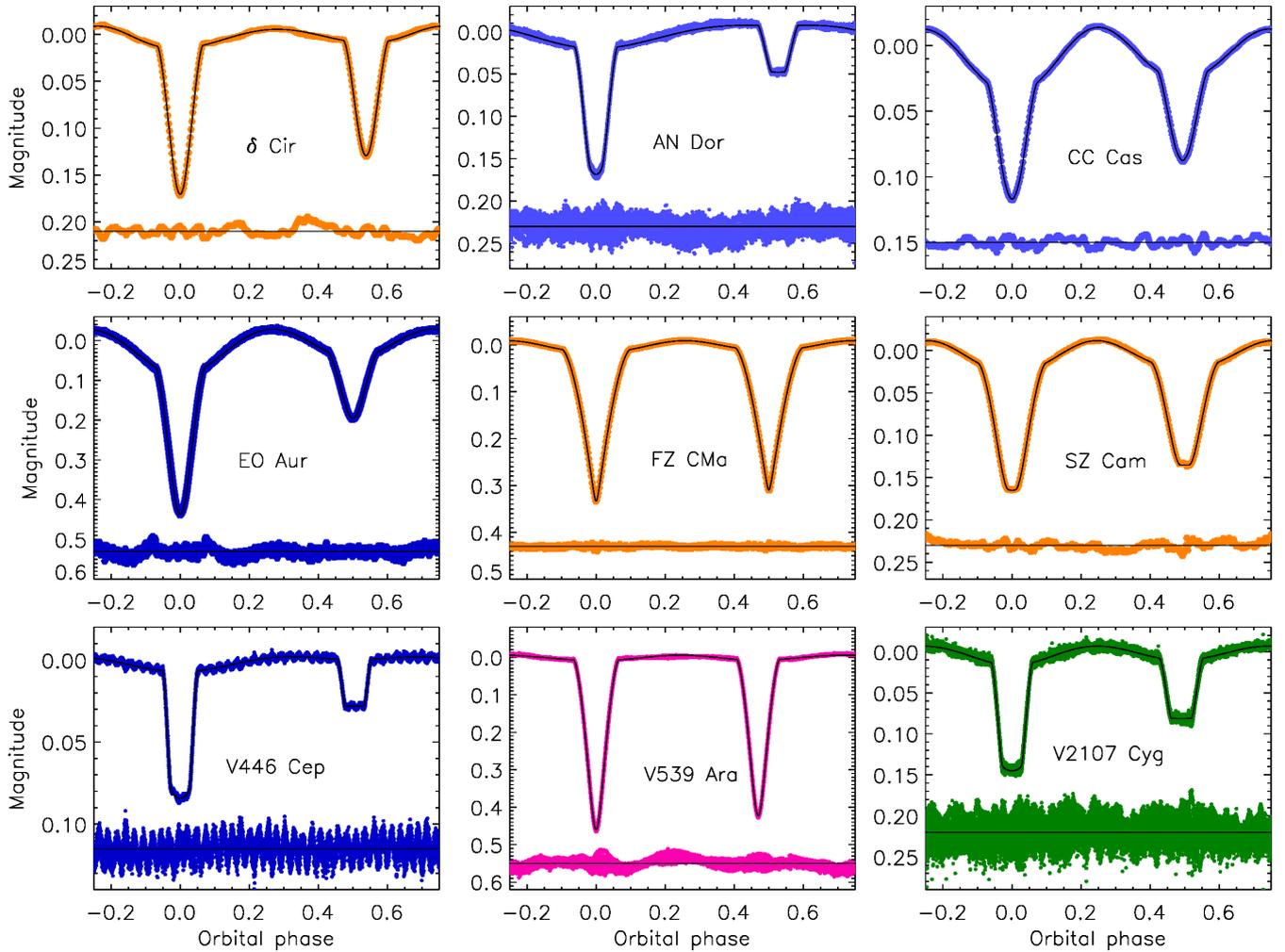}
\caption{\label{fig:phase} Fit to the TESS data for nine of the short-period
systems (labelled). In each case the data are shown after subtraction of the
normalisation polynomials. Those with lots of data were fitted using {\sc jktebop}
and those with only 400 datapoints were fitted using {\sc wd2004}. The best fits
are shown with black lines. The residuals of the fits are shown at the base of each
panel, multiplied by a factor of five to make the features easier to see. The data
shown for AN\,Dor cover only one of the five TESS sectors, in order to decrease
the size of the image file.} \end{figure*}

\begin{figure*} \includegraphics[width=\textwidth]{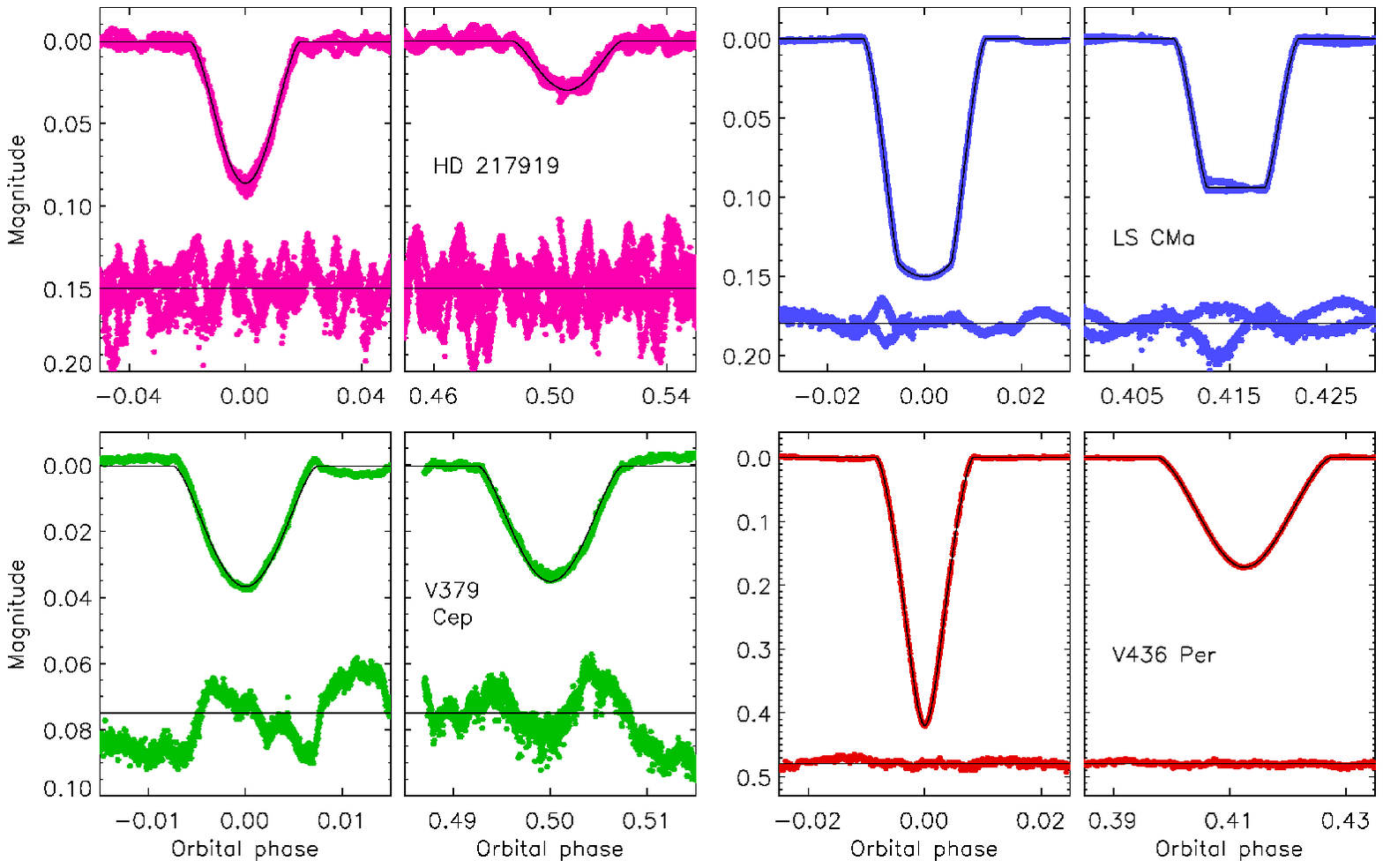}
\caption{\label{fig:ecl} Fit to the TESS data for four of the long-period
systems (labelled). In each case the data are shown after subtraction of the
normalisation polynomials. The {\sc jktebop} best fits are shown with black
lines. The residuals of the fit is shown at the base of each panel, multiplied
by a factor of five to make the features easier to see.} \end{figure*}

%

\subsection{16\,Lacertae $=$ EN\,Lacertae}



16\,Lac is a very bright early-B-type EB showing shallow eclipses and gorgeous $\beta$\,Cephei pulsations. It was discovered to be a spectroscopic binary by \citet{Lee10apj}, to be pulsating by \citet{Walker51pasp}, and to be eclipsing by \citet{Jerzykiewicz80conf}. Star B has not been detected either photometrically or spectroscopically. Spectroscopic observations have been used to investigate the pulsations and to determine the spectroscopic orbit of the primary component \citep{Lehmann+01aa,Aerts+01aa}. Extensive photometry was obtained by \citet{Jerzykiewicz+15mn}, who performed a frequency analysis and a fit to the eclipses. \citet{Jerzykiewicz+15mn} found multiple frequencies attributable to $\beta$\,Cephei pulsations. 16\,Lac has since been observed by TESS in Sector 16, but is not scheduled to be observed again by this satellite.

The TESS data from Sector 16 show clear multi-periodic pulsations with a maximum amplitude of approximately 0.01\,mag, plus two consecutive eclipses of depth 0.04\,mag (Fig.\,\ref{fig:time:1}). We had to remove the requirement of QUALITY $=$ 0 in order to avoid losing datapoints in the second half of the first eclipse; the flagged data appear to be as reliable as the data with QUALITY $=$ 0. The eclipses are grazing and strongly distorted by the pulsations, so we elected to remove the pulsations before fitting the eclipses.

The secondary eclipse is shallow and smaller than the pulsations. It was not detected by \citet{Jerzykiewicz+15mn}, who attributed this to it being very shallow due to the low \Teff\ of star~B. After removal of the pulsations the secondary eclipse is identifiable at the correct orbital phase (0.511 based on the $e$ and $\omega$ from \citealt{Lehmann+01aa}), representing the first direct detection of light from star~B and confirming the interpretation of \citet{Jerzykiewicz+15mn}. We removed the pulsations in two different ways and modelled both light curves with {\sc jktebop}, finding good consistency between the parameters. The orbital shape parameters ($e$ and $\omega$) were fixed at the values for the spectroscopic orbit of star~A from \citet{Lehmann+01aa}. Agreement is good for all parameters except $J$, for which we find 0.133 and 0.230 for the two light curves. Adopting $J=0.18 \pm 0.05$ and $T_{\rm eff,1} = 22\,500$\,K gives $T_{\rm eff,2} = 14700 \pm 1200$\,K. The current TESS data (Fig.\,\ref{fig:time:1}) are not definitive but do at least provide the first detection of the secondary eclipse. From the $r_{\rm B}$, $i$, $K_{\rm A}$ and $P$ in Table\,\ref{tab:lc:1} we can determine the surface gravity of star~B \citep[see][]{Me++07mn}. We find $\log g_{\rm B} = 3.50 \pm 0.04$ which is lower than that of the primary. One explanation for this would be if star~B is still in the pre-main-sequence evolutionary phase.

Owing to the large pulsation amplitude amplitudes of 16\,Lac relative to the eclipse depths, the eclipse profiles are significantly affected by the pulsations which limits our binary modelling. We manually removed the primary eclipses from the TESS light curve of 16\,Lac prior to our pulsation frequency analysis. The summary figure of 16\,Lac is shown in Fig.~A1, including its pre- and post-clipped light curve and labelled amplitude spectrum. Star~A is known to be a multi-periodic $\beta$\,Cephei star, and our analysis of TESS data reveals a dominant pulsation mode frequency of $5.9105 \pm 0.0001$\,d$^{-1}$ with an amplitude of $5.65 \pm 0.03$\,mmag. We detected a total of eight significant independent pulsation mode frequencies which span the frequency range of $1.5 < \nu < 11.5$~d$^{-1}$, which is a typical range for such stars.

\subsection{$\delta$ Circinus}




$\delta$\,Cir is an O-type eclipsing binary that shows SLF variability in its TESS light curve similar to other O-type stars \citep{Bowman+20aa}. No intrinsic variability in either star has been previously reported, likely due to the difficulty of obtaining high-quality photometry of such a bright star. It has a small eccentricity, apsidal motion, and a third component on a wider orbit. \citet{Penny+01apj} discovered the third star in IUE spectra (see also \citealt{Stickland+93obs}) and inferred spectral types of O7\,III-V, O9.5\,V and B0.5\,V for the three components. Their preferred \Teff\ values were $37500 \pm 1500$\,K, $33000 \pm 1000$\,K and $29000 \pm 2000$\,K, respectively. The most recent study was by \citet{Mayer+14aj}, in which references to earlier work can be found. \citeauthor{Mayer+14aj} favoured a spectral type of O8\,IV for star~A. They also interferometrically resolved star~C using the VLTI, finding a magnitude difference of $\Delta H = 1.75$\,mag (no uncertainty quoted) which corresponds to a fractional contribution of approximately 17\% in the $H$-band.

$\delta$\,Cir was observed by TESS in Sector 12 and the light curve shows eclipses of depth 0.17\,mag (primary) and 0.13\,mag (secondary) onto which a much lower-amplitude variability is superimposed (Fig.\,\ref{fig:time:1}). Our fits require an orbital eccentricity that is small but highly significant and in good agreement with previous studies of this system.

Star~A has a large fractional radius so we produced a preliminary fit of the light curve with {\sc jktebop}, phase-folded it, and performed a detailed analysis with {\sc wd2004} (Table\,\ref{tab:lc:wd} and Fig.\,\ref{fig:phase}). Due to the presence of a known third component we included $\ell_3$ as a fitted parameter, obtaining a value of $0.277 \pm 0.023$. This is much larger than the 0.17 expected from the interferometrically-determined $H$-band magnitude difference, suggesting the presence of a fourth component, contaminating light in the large TESS aperture, or imperfect subtraction of background light during the data reduction process. \citealt{Mayer+14aj} noted that some of the observed spectral line profiles were better fitted by four components,  although its hierarchical position within the system is unclear, which supports the larger third light value we find. Future analyses would benefit from additional constraints on $\ell_3$ in the TESS passband, perhaps through further optical or near-IR interferometry.

We determined the physical properties of the system from the results of our WD analysis and the velocity amplitudes from \citet{Penny+01apj}. The velocity amplitude of star~B found by \citet{Mayer+14aj} is significantly larger, but was not supplied with an errorbar so we were do not  use it. We find masses in good agreement with those from \citet{Penny+01apj} but not \citet{Mayer+14aj}, and radii in good agreement with those from \citet{Mayer+14aj}. Although the mass measurements have precisions of 2--3\% their true uncertainty is larger than this; the radius measurements should be reliable.

We detected no significant independent pulsation modes in the residual amplitude spectrum of $\delta$\,Cir, as shown in Fig.~A2. The light curve is dominated by SLF variability, which is typical for stars of spectral type of mid-to-late O. Furthermore, we note that since the system contains more than a single O-type star, the light curve (and amplitude spectrum) dominated by SLF variability has contributions for stars of different mass and age. Hence the flux dilution of each star's variability based on their relative light contributions is an important factor when interpreting SLF variability in multiple systems.

\subsection{$\eta$ Orionis}



$\eta$\,Ori is a system of at least five early-type stars showing multiple types of photometric variability. Components A and B form a visual double separated by 1.8\as, and component C is distant by 114\as\ \citep{Balega+99astl,Mason+01aj}. Component A is itself a spectroscopic triple system comprising an EB \citep{KunzStebbins16paas} with a period of 7.99\,d ($\eta$\,Ori\,Aa,b) orbited by a tertiary with a 9\,yr period ($\eta$\,Ori\,Ac). A photometric variability with a period of 0.301\,d has been found to occur in the AB system, possibly arising from star~Ab \citep{Koch++80baas,BeardsleyZizka80baas}. This periodicity was corrected to 0.432\,d by \citet{WaelkensLampens88aa}. \citet{Demey+96aa} obtained spectroscopic observations and detected LPVs with a period of 0.13\,d that could be ascribed to a non-radial pulsation mode of angular degree $\ell=4$ and azimuthal order $m=-3$. The Aa,b and Ac system has also been spatially resolved using lucky spectroscopy \citep{Maizapellaniz+18aa}. These authors found spectral types of B0.7\,V for Aa and B1.5\,V for B, but were unable to identify Ab in their spectrum. \citet{Maizapellaniz+18aa} also found that component B consists of two stars with high rotational velocities and a large radial velocity (RV) separation. This is consistent with the idea suggested by \citet{Lee+93aspc} that component B is a 0.864\,d contact binary and the shorter-period variation is actually due to the effects of binarity in this object.

$\eta$\,Ori was observed by TESS in sectors 6 and 32 (Fig.\,\ref{fig:time:1}), and is not scheduled to be observed again. Both light curves show eclipses with a period of 7.988\,d and a sinusoidal variability with a 0.43\,d period. The features in the data from sector 32 have a much lower amplitude, implying difficulties in extracting reliable light curves for a star this bright, so we restricted our analysis of the data from sector 6. Here the eclipses are of depth 0.23 and 0.21 mag, and the sinusoidal variability has a period of 0.43207\,d and an amplitude of 0.021\,mag.

Our initial fits to the eclipses in the TESS data showed large residuals due to the shorter-period variation, so we proceeded with fits including a sine term to account for it. This yields a much better result (Table\,\ref{tab:lc:1}), but a more detailed analysis is needed for this object. The ratio of the variability periods is $18.4869 \pm 0.0001$, which is not consistent with any orbital commensurabilities. The shorter-period variability appears to be strictly periodic and also not exactly sinusoidal in shape, consistent with the ellipsoidal effect arising from a close binary. We conclude that the available data are plausibly \reff{explained} as arising from a quintuple star system containing two binary systems: a detached EB with a period of 7.988\,d and one component showing g-mode pulsations, and a non-eclipsing binary with a period of 0.8641\,d and strong ellipsoidal variations.

\subsection{$\lambda$ Scorpii}



$\lambda$\,Sco is a very bright system that has been studied extensively in the past. The inner binary is eclipsing and is composed of a B-star ($10.4 \pm 1.3$\Msun) showing $\beta$\,Cephei pulsations and a 1.6--2.0\Msun\ unevolved MS star; it has an orbital period of 5.953\,d and eccentricity of 0.26. The tertiary component is a B-star ($8.1 \pm 1.0$\Msun) on a much wider orbit of period 2.96\,yr that has been interferometrically resolved. Extensive information and analysis of this system can be found in \citet{Demey+97aa}, \citet{Uytterhoeven+04aa,Uytterhoeven+04aa2} and \citet{Tango+06mn}, and a more recent study can be found in \citet{HandlerSchwarzenbergczerny13aa}.

$\lambda$\,Sco was observed by TESS in Sectors 12 (Fig.\,\ref{fig:time:1}) and 39, but only the first was available at the time of our analysis. We accepted all measurements with finite SAP flux values, irrespective of their QUALITY flag, giving 16\,088 brightness measurements. The TESS light curve exhibits shallow eclipses and strong $\beta$\,Cephei variability. However, the eclipse depths (0.015 and 0.005 mag) are in poor agreement with those in an unpublished light curve (0.04 and 0.01 mag) from the WIRE satellite \citep{BrunttSouthworth08conf} so these data may not be reliable.

We nevertheless proceeded to fit the TESS data with {\sc jktebop} (Table\,\ref{tab:lc:2}). We fixed the third light at 0.54 based on the interferometric measurement at 700\,nm from \citet{Tango+06mn}. Our solution has a much lower eccentricity than in previous studies, and we recommend that a detailed analysis using more extensive data is performed when possible. The results are given in Table\,\ref{tab:lc:1} to only a small number of significant figures. With the $K_{\rm A}$ from \citet{Uytterhoeven+04aa} we were able to calculate the surface gravity of star~B to be 4.44 (log cgs), which is appropriate for a low-mass MS star. This conclusively rules out the possibility that star~B is a white dwarf \citep{Berghofr+00apj} and also disfavours the idea that it is a pre-MS star \citep{Uytterhoeven+04aa}.


Similar to 16\,Lac, $\lambda$\,Sco is a system for which the eclipse profiles are significantly affected by the pulsations which limits our binary modelling. We manually removed the eclipses from the TESS light curve prior to our frequency analysis, with its summary figure shown in Fig.\,A1. The primary of $\lambda$\,Sco is a known multi-periodic $\beta$\,Cephei star, and our analysis of the clipped TESS light curve reveals a dominant pulsation mode frequency of $4.6790 \pm 0.0001$\,d$^{-1}$ and amplitude of $4.02 \pm 0.02$\,mmag, which is consistent in frequency to that detected by \citet{Uytterhoeven+04aa2} using spectroscopy. We detected a total of five significant independent pulsation mode frequencies which span the frequency range of $4 < \nu < 10$~d$^{-1}$. $\lambda$\,Sco was also recently identified as a candidate high-priority target for the upcoming ESA/KU Leuven CubeSpec space mission, which will assemble high-cadence and high resolution optical spectroscopy of massive stars, because of its high amplitude $\beta$\,Cephei pulsations \citep{Bowman+22aa}.

\subsection{$\mu$ Eridani}



$\mu$\,Eri is another bright system with an extensive observational history, composed of an SPB star and a much smaller and less massive star in a 7.359\,d orbit with an eccentricity of 0.34. It has been known to be a spectroscopic binary for over a century \citep{FrostAdams10sci,Frost++26apj,Hill69pdao}, and was discovered to be a pulsating variable by \citet{Handler+04mn} and eclipsing by \citet{Jerzykiewicz+05mn}. \citet{Jerzykiewicz+13mn} presented a detailed analysis of the system based on extensive spectroscopy and 12\,d of observations from the MOST satellite. They extracted pulsation frequencies, fitted the eclipses, and derived a new single-lined spectroscopic orbit.

$\mu$\,Eri was observed by TESS in Sectors 5 and 32, totalling 52\,d of coverage including observations of seven eclipses (Fig.\,\ref{fig:time:1}). The pulsations strongly affect the eclipse shapes but unfortunately are not easy to remove. We therefore modelled the two sectors of data with {\sc jktebop} with the pulsations still present, and give only indicative parameters in Table\,\ref{tab:lc:2}. We fixed $e$ and $\omega$ at the values given by \citet{Jerzykiewicz+13mn} and chose an indicative surface brightness ratio of 0.15. We expect that a more detailed analysis of these data could lead to a more robust model of the system.

Owing to the significant gap between the two available short-cadence sectors of $\mu$\,Eri, we analysed the pulsational variability of both sectors independently. After manually removing the eclipses from the light curves, we did not find any significant frequencies following our S/N $\geqslant$ 5 significance criterion, despite $\mu$\,Eri being a known g-mode pulsator \citep{Handler+04mn,Jerzykiewicz+05mn,Jerzykiewicz+13mn}.  This is not surprising since in our analysis we restricted ourselves to S/N $\geqslant$ 5\reff{, and the resolution of the currently available TESS light curves is insufficient to reliably extract frequencies from a multiperiodic g-mode pulsator,} whilst previous studies have pushed down to as low as S/N $\geqslant$ 3. The low-frequency g-mode regime of $\mu$\,Eri, as can been seen in Fig.\,A1, is dense and contains multiple unresolved frequencies. These unresolved frequencies together increase the local noise level resulting in the highest amplitude peaks having S/N~$<$~5. The analysis of both pulsations and eclipses in $\mu$\,Eri will greatly benefit from further spectroscopic monitoring.

\subsection{AN Doradus}


AN\,Dor shows total eclipses with the secondary (0.05\,mag) much shallower than the primary (0.16\,mag), plus a strong reflection effect and clear $\beta$\,Cephei pulsations. Little is known about this object: it was discovered to be eclipsing using {\it Hipparcos} satellite data and given its designation in the General Catalogue of Variable Stars by \citet{Kazarovets+99ibvs}. \citet{HoukCowley75book} gave its spectral type as B2/3\,V. \citet{PercyAuyong00ibvs} included it in a short list of variable B-stars in EBs, with the comment that ``the short-term variability is \ldots uncertain''. No other study of it has been published, to our knowledge.

AN\,Dor has been observed extensively by TESS. It was observed in short cadence in two sets of five consecutive sectors (2--6 and 29--33), in long cadence in sectors 9, 12 and 13, and again in short cadence in sectors 36 and 39. To obtain a preliminary characterisation of the system we have analysed the light curve from sectors 29--33 (see Fig.\,\ref{fig:time:1}) as these are sufficient for our purposes and of better quality that that from sectors 2--6. We find a good fit using {\sc jktebop} (Fig.\,\ref{fig:phase}) although star~A is formally too deformed for this code to be reliable (see Table\,\ref{tab:lc:2}). Despite the short orbital period of 2.03\,d the system has an eccentric orbit. Star~B is smaller and much fainter than star~A; the light ratio of 0.7\% means extensive effort will be needed to obtain RVs for it.

For the \refff{frequency} analysis we again restricted ourselves to analysing sectors 29--33, as these data cover a long time interval and are of high quality. There are many significant frequencies in the residual amplitude spectrum of AN\,Dor that coincide with integer multiples of the orbital frequency. These are shown as dashed red lines in the summary figure of AN\,Dor in Fig.\,A2. We interpret the majority of these frequencies to be the result of an imperfect binary model. On the other hand, we also detect many significant independent frequencies which are indicated by green lines in Fig.\,A2. Given the frequency range of its variability, spanning between 1 and 16\,d$^{-1}$ with its dominant frequency range between 1 and 4~d$^{-1}$, and spectral type, we conclude that AN\,Dor is a $\beta$\,Cephei \refff{and/or SPB pulsator. Further observations and analysis are needed to refine this conclusion.}

\subsection{CC Cassiopeiae}




CC\,Cas is an EB containing two close but detached stars of masses 23\Msun\ and 10\Msun\ on a 3.37\,d orbit. It was discovered to be an SB2 by \citet{Pearce27pdao}, to be eclipsing by \citet{GuthnickPrager30an}, and to show night-to-night variability by \citet{Polushina88pz}. The most detailed study of the system was published by \citet{Hill+94aa}, who measured the masses, radii, \Teff\ values and spectral types of the component stars. A more recent spectroscopic analysis by \citet{Gorda13astbu} returned significantly larger masses for the two stars than in previous works, suggesting that more extensive study of this spectroscopically difficult system is needed.

CC\,Cas was observed by TESS in Sectors 18 and 19 (Fig.\,\ref{fig:time:2}) and the appearance of the light curve in the two sectors is very similar. The large fractional radii of star~A meant we had to perform an initial {\sc jktebop} analysis to obtain the orbital ephemeris, phase-bin the data, and then fit it with {\sc wd2004} (Fig.\,\ref{fig:phase}). The results of this work are given in Table\,\ref{tab:lc:wd}. The fractional radii are measured to precisions of 1.2\% and 2.5\%. The orbital inclination, $65.44 \pm 0.21^\circ$, is one of the lowest known for an EB and is possible because of the large sizes of the stars relative to the orbit. It is also significantly lower than found in previous work (e.g.\ \citealt{Hill+94aa} found $i = 69.6 \pm 0.4^\circ$) -- this may have occurred due to the low quality of the photometry available before TESS or alternatively it might indicate presence of dynamical effects such as those as seen in VV\,Ori by \citet{Me++21mn}.

The light curve solution requires a small but highly significant orbital eccentricity of $e = 0.0099 \pm 0.0013$; solutions assuming a circular orbit cannot correctly reproduce the shapes of the ingress and egress of the eclipses. Previous studies of CC\,Cas (e.g.\ \citealt{Hill+94aa} and \citealt{Gorda13astbu}) have assumed a circular orbit due to lack of evidence for eccentricity, but the TESS light curve is inconsistent with that assumption. We determined the physical properties of the system using our photometric analysis and the spectroscopic results from \citet{Gorda13astbu}, which appear to be based on the highest-quality spectra. The masses are measured to better than 4\% precision and the radii to better than 2\%. This system is therefore a good candidate for measuring the properties of a 23\Msun\ star to high precision, although extensive high-quality spectroscopy and a sophisticated analysis will be needed.

We detect no significant independent pulsation modes in the residual amplitude spectrum of CC\,Cas, as shown in Fig.~A2. The light curve is dominated by SLF variability, which is consistent with the recent result that main-sequence O-type stars have predominantly SLF variability rather than multiple high-amplitude heat-driven modes \citep{Bowman+20aa}.

\subsection{EO Aurigae}




EO\,Aur is an EB containing two early-B stars in a 4.07\,d orbit. The system was discovered to be a spectroscopic binary by \citet{Pearce43paas} and to be eclipsing by \citet{Gaposchkin43pasp}. It has proved to be spectroscopically intractable due to the large line broadening of the components \citep{Popper78apj,Burkholder++97apj} so the masses are not known with any certainty; future analysis using methods such as spectral disentangling may meet with more success \citep[e.g.][]{PavlovskiHensberge05aa,Pavlovski++18mn}. Light curves and radius measurements have been presented by \citet{Schneller63an} and \citet{Hartigan81jaavso}.

EO\,Aur was observed in TESS Sector 19 (Fig.\,\ref{fig:time:2}). Our {\sc jktebop} fit to these data is shown in Fig.\,\ref{fig:phase}. We get a reasonable but not a good fit, failing in particular to match the data around the times of first and fourth contact for the primary eclipse. Star~A has a fractional radius too large for {\sc jktebop} (Table\,\ref{tab:lc:2}) and this is likely the reason for our imperfect fit. The system is in need of extensive new observations and analysis for its properties to be established reliably.

The residual amplitude spectrum of EO\,Aur contains four significant frequencies which coincide with integer multiples of the orbital frequency, but also a single additional independent frequency at $12.1466 \pm 0.0005$\,d$^{-1}$. Therefore we classify EO\,Aur as containing a $\beta$\,Cephei pulsator. The summary figure of EO\,Aur is shown in Fig.\,A2. We surmise, similarly to many stars in our sample, that EO\,Aur is a multi-periodic $\beta$\,Cephei pulsator, but the analysis of only a single available TESS sector does not reveal low amplitude pulsation modes.

\subsection{FZ Canis Majoris}




FZ\,CMa was discovered to show double spectral lines with variable RV by \citet{Neubauer43apj} and to be eclipsing on a 1.27\,d period by \citet{MoffatVogt74aaa}. A detailed analysis of the system was presented by \citet{Moffat+83aa}. They found excess scatter in their light curves and attributed this to intrinsic variability of the system. They also found a large third light and a light-time effect indicative of the presence of a tertiary component of large mass and an orbit with a period of 1.47\,yr. The system has been spectrally classified as B2\,Vn by \citet{Claria74aj}, as B2.5\,IV-V by \citet{MoffatVogt74aaa}, and as B2\,IVn by \citet{Herbst+78apj}. \citet{Moffat+83aa} obtained eight spectra which showed lines of the eclipsing stars and used these to determine preliminary masses for them. They did not spectroscopically detect the tertiary despite its large mass, which suggests it is either rotating very quickly or instead could itself be binary \citep{Chambliss92pasp}. A detailed spectroscopic study of this system is needed in order to characterise it properly.

FZ\,CMa was observed by TESS in sectors 7 and 33, and both light curves show clear eclipses and intrinsic variability (Fig.\,\ref{fig:time:2}). Due to the known light-time effect we modelled these separately using {\sc wd2004}. The eclipses are shallow and V-shaped, and can only be fitted with a large amount of third light, as already found by \citet{Moffat+83aa}. The amount of third light is very well determined as $\ell_3 = 0.467 \pm 0.007$, i.e.\ it is almost half of the total light of the system. We are able to obtain a good but not perfect fit to the TESS data (Fig.\,\ref{fig:phase}), and determine the fractional radii to precisions of 0.2\% for the primary and 1.0\% for the secondary (Table\,\ref{tab:lc:wd}). To obtain our solution we adopted a \Teff\ for star~A of 22\,000\,K \citep{Moffat+83aa} and arbitrarily assigned an uncertainty of 2000\,K to this value. The physical properties are given in Table\,\ref{tab:lc:wd} and suffer from large uncertainties in the measured velocity amplitudes.

We analysed the sector 7 and 33 data for FZ\,CMa separately, finding that they ultimately yield similar pulsation mode frequencies. FZ\,CMa shows a total of four significant pulsation mode frequencies between $2 < \nu < 8$~d$^{-1}$ in sector 7, classifying it as a $\beta$\,Cephei \refff{and/or SPB} star. \refff{The masses of the two stars (Table\,\ref{tab:lc:wd}) are quite low both for$\beta$\,Cephei pulsations and for their spectral types, which has (at least) two plausible explanations. First, line blending due to the fast rotation of the stars causes their masses to be underestimated \citep[e.g.][]{Andersen75aa}. Second, the pulsational signature may arise from the third component, which contributes a large fraction of the total light, rather than one (or both) of the eclipsing stars. In both cases the system would benefit from a detailed spectroscopic analysis to determine precise masses and \Teff\ values for the component stars.} The summary figures for both sectors are shown in Fig.\,A2.

\subsection{HD 217919}




HD~217919 is unique in this work as it was not previously known to be an EB, but was instead picked up by a colleague (Dr.\ P.\ Maxted) whilst browsing the TESS database. It is a known spectroscopic binary, for which \citet{Garmany72aj} presented a single-lined orbit with a period of 17.04\,d and an eccentricity of 0.26; he also noted that the lines appeared doubled on three of the plates. The TESS light curve shows lovely $\beta$\,Cephei pulsations overlaid on obvious but shallow eclipses of depth 0.09\,mag (primary) and 0.03\,mag (secondary)\reff{, representing the first detection of eclipses and pulsations in this system.}

HD~217919 has been observed by TESS in three sectors: 17, 18 and 24 (Fig.\,\ref{fig:time:2}). The phasing of the observations is such that six secondary but only three primary eclipses were observed. The stars are well-detached so are suitable for analysis with {\sc jktebop}. The time interval covered by the TESS data is long enough to get a precise linear ephemeris. In Fig.\,\ref{fig:phase} we show the best fit and in Table\,\ref{tab:lc:2} we give the fitted parameters from our eclipse analysis. We find a solution with a period of 16.2\,d, a small eccentricity, and a significant third light. It is therefore possible that there is a tertiary component that is responsible for the pulsations, rather than one of the eclipsing components. The presence of a bright third star will also make spectroscopic analysis of this system difficult. The solution is unstable in that separate fits of individual sectors give very different results, so we give only an indicative solution: the best fit to all data but to only a few significant figures.

We performed a frequency analysis of the two segments of TESS data available for HD~217919 separately, although they differ somewhat in quality and the resultant frequency lists. This revealed a rather dense spectrum of significant pulsation modes that span between $1.5 < \nu < 10.5$~d$^{-1}$. The apparent groups of the frequencies is reminiscent of high-radial order gravity modes seens in non-linear SPB pulsators (see e.g.\ \citealt{Kurtz+15mn}), which can be explained at least in part by combination frequencies. However, such a phenomenon has not been seen for low-radial order p-modes in $\beta$\,Cephei stars, assuming that these pulsation modes are intrinsic to the primary. If they are intrinsic to the secondary or \reff{tertiary}, then the companions could be SPB stars. Given the frequency range and spectral types of the system, we tentatively classify these as $\beta$\,Cephei \refff{/ SPB} pulsations. They may plausibly arise in any of the three components of the system.

\subsection{HQ Canis Majoris}




HQ\,CMa was found to show variable RV by \citet{BuscombeMorris60mn} and to be eclipsing by \citet{JerzykiewiczSterken77aca}. \citet{Sterken+85aas} presented the only known dedicated analysis of this object, establishing its orbital period as 24.6033\,d. although they were not able to observe a single eclipse in its entirety.

HQ\,CMa has been observed by TESS in three sectors: 7, 34  and 35 (Fig.\,\ref{fig:time:2}). Only one eclipse is visible in these data, in sector 34 and of depth 0.022\,mag, but variability consistent with SPB and/or $\beta$\,Cephei pulsations is clearly discernable. Upon analysis with {\sc jktebop} we immediately discovered the period from \citet{Sterken+85aas} is incorrect because the data 24.6\,d after the observed eclipse does not have an eclipse. A nearby gap in the data allows periods of 21.2 to 22.6\,d; values of 34.5\,d or longer would also be consistent with the TESS light curve. Faced with a system showing a single shallow partial eclipse and an unknown orbital period, it is a thankless task to deduce its properties so we have not attempted to do so. The shallowness of the eclipse can easily be matched using a grazing eclipse configuration, but it is also possible that there is a strong third light in the system in which case it is possible that the pulsations do not arise from either of the stars in the EB.

We analysed sectors 33 and 34 of the TESS data of HQ~CMa in search for significant pulsations, blind to in which star they may originate. As shown in the summary figures in Fig.\,A1, there is evidence of both low-frequency g-modes indicative of an SPB star and high-frequency p-modes indicative of a $\beta$\,Cephei star. The amplitudes of the $\beta$\,Cephei pulsations are quite small and not all frequencies are significant in both sectors. Similarly, the presence of multiple unresolved frequencies in the low-frequency g-mode regime means that only a single g-mode frequency has S/N~$\geq$~5 in sector 33. Since the primary has a spectral type of B3\,V, it is in the mass regime that allows for both p- and g-mode frequencies to be excited by the $\kappa$ mechanism during the main sequence \citep{Walczak+15aa,SzewczukDaszynska17mn}. We conclude that HQ\,CMa contains either a candidate SPB/$\beta$\,Cephei pulsator, or the $\beta$\,Cephei pulsations originate in a contaminating source given that the secondary likely has a later spectral type that the primary, and hence is not massive enough to host $\beta$\,Cephei pulsations.

\subsection{LS Canis Majoris}



Despite its brightness, almost nothing was previously known for LS\,CMa. Its variability was found using the {\it Hipparcos} satellite and it was attributed this name and the ``E:'' designation in the General Catalogue of Variable Stars \citep{Kazarovets+99ibvs}. Its RV is listed as 6\kms\ in the General Catalogue of Stellar Radial Velocities \citep{Wilson53gcvr} and a rotational velocity of 17\kms\ was given by \citet{Abt++02apj}.

LS\,CMa was observed with TESS in two sets of two consecutive sectors: 6 and 7, and 33 and 34 (Fig.\,\ref{fig:time:2}). The light curve shows clear variability aside from the eclipses. Four eclipses were observed: a secondary then part of the next primary in sectors 6 and 7, and a primary and immediately following secondary in sectors 33 and 34. The primary eclipse is annular and the secondary eclipse is total, indicating that the primary star is significantly hotter and larger than the secondary. Due to the order the eclipses were observed in, and despite the 730\,d time interval between the two observed primaries (and also secondaries) it is straightforward to establish the orbital period of the system as 70\,d. Using {\sc jktebop} we refined this and the remaining photometric parameters to the values given in Table\,\ref{tab:lc:1}. The best fit to the eclipses is shown in Fig.\,\ref{fig:ecl}.


Frequency analysis of sectors 33 and 34 of LS\,CMa reveals a few significant frequencies that coincide with integer multiples of the orbital frequency. We also detect a single independent frequency in sector 33, of frequency $\nu = 1.61381 \pm 0.00006$~d$^{-1}$ and amplitude of $0.768 \pm 0.004$~mmag. This frequency is not significantly detected in sector 34. LS\,CMa is therefore not a convincing pulsating EB system based on the data currently available. The summary figures for the frequency analysis of LS\,CMa are shown in Fig.\,A3.

\subsection{SZ Camelopardalis}




SZ\,Cam is a triple system with an inner orbital period of 2.70\,d and a long observational history, having been discovered to be a spectroscopic binary by \citet{Plaskett24pdao} and to be eclipsing by \citet{GuthnickPrager30an}. Extensive investigations have been published by \citet{Wesselink41anlei,Mayer+10aa} and \citet{Tamajo+12aa}. There is a tertiary star that orbits the EB every approximately 55\,yr. \citet{Tamajo+12aa} found that the third component is itself an SB1 with a period of 2.80\,d, meaning that SZ\,Cam is a quadruple system. They also detected $\beta$\,Cephei pulsations in the system with a period of 0.333\,d and tentatively attributed it to one of the non-eclipsing stars.

SZ\,Cam was observed by TESS in sector 19 (Fig.\,\ref{fig:time:3}). We fitted these data with {\sc wd2004} assuming a circular orbit and allowing for third light (Fig.\,\ref{fig:phase}). The amount of third light we obtain is much greater than found by \citet{Tamajo+12aa}, which can be explained by the large size of the apertures used to extract the light curve from the TESS data. However, many other parameters show a significant difference between our values (Table\,\ref{tab:lc:wd}) and those found by \citet{Tamajo+12aa}. This includes the orbital inclination, which might have changed due to dynamical effects, and the \Teff\ of star~B which is inconsistent with the spectroscopic value from \citet{Tamajo+12aa}. We deduced a value of the mass ratio from the TESS light curve: it is somewhat smaller than that obtained by \citet{Tamajo+12aa} from the spectroscopic orbits, and better matches values from earlier studies \citep[see table\,1 in][]{Tamajo+12aa}. In calculating the physical properties of the system we favoured the spectroscopic mass ratio and \Teff\ valus from \citet{Tamajo+12aa} over our own determinations. The measured radii are much less certain than the values from \citet{Tamajo+12aa} despite the use of the TESS data. Based on these issues, we conclude that the properties of SZ\,Cam are not as well established as suggested by previous work. A new analysis of this spectroscopically-difficult system is needed.

We detect no significant independent pulsation modes in the residual amplitude spectrum of SZ\,Cam, as shown in Fig.\,A3. The light curve is dominated by SLF variability. This is not surprising given that the majority of the light contribution will be from the O9\,IV primary and O-type stars are known to be dominated by SLF variability \citep{Bowman+20aa}. We do not detect the $\beta$\,Cephei pulsation mode frequency of \citet{Tamajo+12aa} at significant amplitude in the TESS data. This is probably because of flux dilution of the pulsating star by the O-type primary, and the high fraction of contamination from nearby stars for SZ\,Cam given that it is a member of a young cluster \citep{Tamajo+12aa} and TESS pixels subtend a large angular size.

\subsection{V379 Cephei}



V379\,Cep was found to exhibit variable RV by \citet{Adams++24pasp} and to be eclipsing by \citet{Jerzykiewicz93aas}; its orbital period of 99\,d took some effort to establish \citep{Clayton96pasp,Gordon+98aj}. \citet{Harmanec+07aa} found it to be a hierarchical quadrule system composed of two binary systems (the non-eclipsing one having a period of 158.7\,d) orbiting each other every 7900\,d.

V379\,Cep was observed by TESS in sectors 15--17 and will be observed again in sector 55. Two eclipses are visible, one in sector 15 and one in sector 17, separated by 49.9\,d (Fig.\,\ref{fig:time:3}). Under the assumption that these are one primary and one secondary eclipse, we fitted the light curve with {\sc jktebop} using a fixed orbital period of 99.7658\,d \citep{Harmanec+07aa}. Our solution (Fig.\,\ref{fig:ecl}) is surprisingly consistent with a circular orbit, and unsurprisingly requires a large and poorly-constrained amount of third light to match the data. We adopted $\ell_3 = 0.50$ to produce an indicative solution of the TESS data, which is given in Table\,\ref{tab:lc:2}. This is lower than the value of $\ell_3 = 0.7$ adopted by \citet{Harmanec+07aa}, but returns (slightly) more plausible system properties from the TESS light curve. Further work on this object is needed for its properties to be reliably established.

We detect no significant independent pulsation modes in the residual amplitude spectrum of V379\,Cep, as shown in Fig.~A3. The light curve is dominated by SLF variability.

\subsection{V436 Persei}



\begin{figure} \includegraphics[width=\columnwidth]{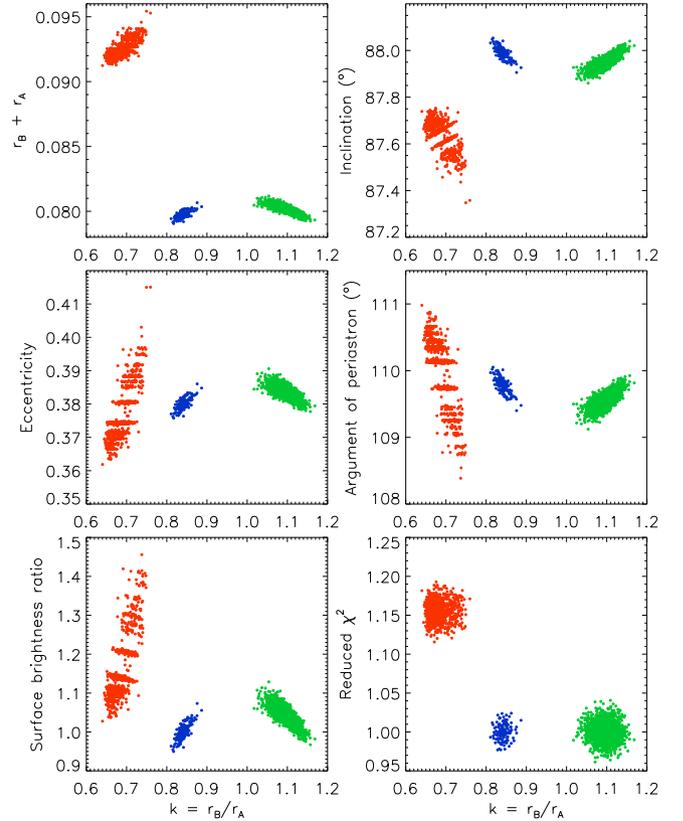}
\caption{\label{fig:v436per:mc} Monte Carlo scatter plots for V436\,Per. Three local minima in parameter
space are visible in the ratio of the radii. These have been represented in different colours for clarity.
The preferred minimum is the one in green and is the group with the largest value of $k$.} \end{figure}

V436\,Per was found to be a spectroscopic binary by \citet{Adams12apj} and to be eclipsing by \citet{Kurtz77pasp}. \citet{Harmanec+97aa} have summarised the observational history of the system, and also detected LPVs in their high-resolution spectra. These authors subsequently revisited the system \citep{Janik+03aa}, were unable to confirm the LPVs, and established the orbital elements to high precision using spectral disentangling.

Observations of V436\,Per were obtained by TESS in sector 18 (Fig.\,\ref{fig:time:3}), and the light curve fortuitously contains one primary and one secondary eclipse (Fig.\,\ref{fig:phase}). To constrain the orbital period of the system we adopted a time of minimum of HJD $2443562.861 \pm 0.020$ where the value comes from \citet{Janik+03aa} and the errorbar from \citet{Harmanec+97aa}. The $e$ and $\omega$ of the system are of such a value that there exists three regions in parameter space that provide a similar fit. This is illustrated in Fig.\,\ref{fig:v436per:mc}.

We can reject the local minimum with $k = 0.69 \pm 0.03$ on statistical grounds, as it corresponds to a significantly worse fit (higher reduced $\chi^2$). On closer inspection this result occurs only for unphysical values of the limb darkening coefficients. We can also reject the local minimum with $k = 0.84 \pm 0.02$ on astrophysical grounds, as it corresponds to an inverted mass--radius relationship for this binary system composed of two main-sequence stars. We therefore base our results on the third minimum ($k = 1.10 \pm 0.03$), which also is fully consistent with the $e$, $\omega$ and \Teff\ values from the extensive spectroscopic analysis of this system by \citet{Janik+03aa}. With our results plus the velocity amplitudes from \citet{Janik+03aa} we establish physical properties of the system to high precision for the first time (Table\,\ref{tab:lc:1}). A confirmation of these results could come from a direct measurement of a spectroscopic light ratio, as this differs significantly between the three local minima.

We detect no significant independent pulsation modes in the residual amplitude spectrum of V436\,Per, as shown in Fig.~A3. The light curve is dominated by SLF variability, but additional photometry would be useful in distinguishing SLF variability from independent pulsation modes that are possibly present but insignificant in the current data.

\subsection{V446 Cephei}




V446\,Cep was discovered to be eclipsing based on data from the \textit{Hipparcos} satellite \citep{Kazarovets+99ibvs}, with a period of 3.81\,d. \citet{Cakirli+14} published the only dedicated analysis of the system, based on the \textit{Hipparcos} light curve and 15 medium-resolution spectra. They assigned spectral types of B1\,V and B9\,V, and determined the masses (18 and 2.6\Msun) and radii (8.3 and 2.1\Rsun) of the stars.

TESS observed V446\,Cep in sectors 16, 17 and 24. The light curve shows annular primary eclipses of depth 0.08\,mag, total secondary eclipses of depth 0.03\,mag, and multiperiodic pulsations of a few mmag amplitude (Fig.\,\ref{fig:time:3}). Our {\sc jktebop} analysis required a small orbital eccentricity to fit the data properly, but no third light was needed. We were able to get a good fit to the data (Fig.\,\ref{fig:phase}) and evaluated the uncertainties of the resulting parameters using Monte Carlo and residual-permutation algorithms \citep{Me08mn}.

The surface brightness ratio of the two stars is 0.48 so their \Teff\ ratio should be approximately $0.48^{1/4}=0.83$. This does not agree with the determinations by \citet{Cakirli+14}, $26600 \pm 1000$\,K and $11900 \pm 1050$\,K, which have a ratio of $0.45 \pm 0.05$. Given this discrepancy, and the preliminary nature of the published analysis, their spectroscopic results are questionable. We have therefore not determined the physical properties of the system, in favour of deferring this until more extensive spectroscopy becomes available. If the \Teff\ of star~A from \citet{Cakirli+14} is reliable, then our $J$ gives a \Teff\ of star~B of approximately 22\,000\,K (Table\,\ref{tab:lc:1}).

We performed a \refff{frequency} analysis of sectors 16--17 and 24 of V446\,Cep separately owing to their temporal separation and difference in data quality. Most noteworthy is the presence of a high-amplitude frequency in both residual amplitude spectra which coincides with an orbital harmonic, as shown in the summary figures in Fig.\,A3. Specifically, $10.24372 \pm 0.00006$\,d$^{-1}$ with an amplitude of $1.262 \pm 0.007$\,mmag in sectors 16--17 and $10.2437 \pm 0.0002$\,d$^{-1}$ with an amplitude of $1.28 \pm 0.01$\,mmag in sector 24. There are also several independent pulsation modes spanning $2 < \nu < 12$~d$^{-1}$. Therefore, we conclude that V446\,Cep is a candidate system for showing TEOs in addition to $\beta$\,Cephei pulsations. On the other hand, if the secondary is sufficiently evolved it could be a $\delta$\,Scuti pulsator, but we deem this a less favourable solution given the known properties of the system.

\subsection{V539 Arae}




V539\,Ara consists of B3\,V and B4\,V stars in a 3.17\,d orbit. \citet{Neubauer30licob} found it to be SB2 and \citet{Strohmeier64ibvs3} discovered eclipses. The most detailed work on this object comes from the Copenhagen group: \citet{Andersen83aa} determined a spectroscopic orbit based on extensive photographic spectroscopy, \citet{Clausen+96aas} presented extensive $uvby$ photometry, and \citet{Clausen96aa} measured the masses and radii of the component stars to high precision. The system shows intrinsic variability \citep{Knipe71aa}, which \citet{Clausen96aa} ascribed to SPB pulsations in the secondary star and identified three possible frequencies. V539\,Ara also undergoes apsidal motion \citep{Andersen83aa} with an apsidal period of $162 \pm 8$\,yr. There is a tertiary companion with a period of $42.3 \pm 0.8$\,yr \citep{WolfZejda05aa} and a wider companion at 12.3\,arcsec that is fainter than the EB by approximately 3.6\,mag in the \textit{Gaia} $G$, BP and RP passbands.

V539\,Ara was observed by TESS in sectors 13 and 39 (Fig.\,\ref{fig:time:3}). For our {\sc jktebop} analysis we analysed these separately and did not consider photometry from other sources, in order to avoid issues with the apsidal motion. We allowed for an eccentric orbit and third light, and calculated uncertainties using the Monte Carlo and residual-permutation algorithms (Fig.\,\ref{fig:phase}). The results from the two sectors are in very good agreement so were combined according to their weighted means. The residual-permutation errorbars are larger than the Monte Carlo errorbars, which can be attributed to the variability in the light curve. Our results establish the radii of the stars to precisions of 0.4\% (star~A) and 1\% (star~B; Table\,\ref{tab:lc:1}), and are in good agreement with those from \citet{Clausen96aa}.

We detect no significant independent pulsation modes in the residual amplitude spectrum of V539\,Ara, as shown in Fig.~A3, hence it is dominated by SLF variability. Similarly to SZ\,Cam, the difference in our classification of SLF variability and the SPB classification of  \citet{Clausen96aa} is likely because of flux dilution of the pulsating star by the non-pulsating star, and possible contamination from the large TESS pixels. If additional light curve data become available, it is likely that the independent g-mode frequencies of \citet{Clausen96aa} would become extractable and significant.

\subsection{V2107 Cygni}


V2107\,Cyg was observed spectroscopically by \citet{Mercier57jo}, who found it to be SB1 with a small eccentricity. It was subsequently detected to be eclipsing from \textit{Hipparcos} observations \citep{Kazarovets+99ibvs}. The only detailed study published so far is that of \citet{Bakis+14aj}, who presented medium-resolution spectroscopy and extensive $BVR$ photometry. They established the physical properties of the stars and detected LPVs, attributing this to $\beta$\,Cephei pulsations in the primary star.

V2107\,Cyg has been observed by TESS in sectors 14, 15 and 41 (Fig.\,\ref{fig:time:3}), and will also be observed in sectors 54 and 55. These show clear $\beta$\,Cephei pulsations superimposed on total and annular eclipses. Our {\sc jktebop} fit (Fig.\,\ref{fig:phase}) included an eccentric orbit but not third light, as this is not well determined by the data. The primary star is tidally deformed beyond the limits of applicability of {\sc jktebop} \refff{\citep{Kopal59book,NorthZahn04newar}}, and the velocity amplitude of the secondary star is uncertain \citep{Bakis+14aj} so we give only approximate values for the properties of the components (Table\,\ref{tab:lc:2}).

Our frequency analysis of the residual light curve of V2107\,Cyg reveals a multiperiodic pulsator with a frequency range spanning $0.37 < \nu < 9.21$\,d$^{-1}$. Combined with the spectral type of B1\,III, this confirms the pulsator class identification of $\beta$\,Cephei\refff{/SPB}. We find no \refff{obvious} regular structure or patterns in the frequency spectrum of observed pulsations indicative of rotational splittings or tidally perturbed modes, which is interesting given its significantly distorted structure \citep[see e.g.][]{Me+20mn,Me++21mn}. We conclude that the V2107\,Cyg is a valuable system for follow-up study. We are in the process of obtaining new high-quality spectroscopy and will perform a detailed analysis of the system in due course.



\section{Summary and discussion}

We present the analysis of TESS data for 18 eclipsing binary systems containing pulsating high-mass stars. Of these, \refff{six are definite and eight are candidate} $\beta$\,Cephei pulsators, \refff{eight are possible} SPB pulsators and \reff{seven} have light curves and amplitude spectra dominated by SLF variability. Most of the pulsation detections are new, and significantly increase the number of EBs known to contain stars with these types of pulsations. We fitted the light curves with two EB models to determine the properties of the systems and remove the effects of binarity from the TESS data. These residual light curves were then subjected to frequency analysis to classify the pulsations and measure pulsation frequencies and amplitudes. Future work, guided by additional spectroscopy and TESS photometry, will aid in identifying the geometries of the identified pulsation mode frequencies, which is typically a prerequisite for forward asteroseismic modelling.

Spectroscopic orbits are available in the literature for several of these EBs, and in these cases the full physical properties of the system were calculated based on the literature spectroscopy and the TESS data. We determined precise physical properties for five systems ($\delta$\,Cir, CC\,Cas, SZ\,Cam V436\,Per and V539\,Ara) and preliminary physical properties for four more. Many of the objects studied in this work show lovely eclipses and pulsations and are promising candidates for detailed study. We have already begun a spectroscopic survey of the best candidates and will present results once we have them.

The study of pulsating high-mass stars in EBs is an important new avenue for constraining the physics of these objects. In particular, the amount and shape of the interior mixing profiles of massive stars are unconstrained and there is ever-growing evidence that current stellar structure models underpredict the mixing in massive EBs \citep{Tkachenko+20aa}. When coupled with asteroseismology, the conclusion of needing extra mixing is strengthened further \citep{Johnston21aa}.

Another important constraint that pulsations within massive binaries can provide are on the impact of tides on stellar structure. In the case of short-period, near-circular binaries, the effect of tides can modify the equilibrium structure such that this is detected in the pulsation frequency spectrum \citep{Me+20mn,Me++21mn}. There are few such detections of so-called tidally-perturbed pulsation modes in massive binaries. Therefore, a larger survey of constraining the impact of tides on binary star evolution theory probed by pulsations is needed.

Finally, an important constraint that pulsating massive binaries may provide is on the excitation mechanism(s) of pulsation modes. It is known that the heat-engine mechanism is strongly dependent on the opacity and rotation of a star \citep[e.g.][]{SzewczukDaszynska17mn}. However, it is often difficult to determine an accurate mass and age of a massive star from spectroscopy or evolutionary models alone. The dynamical masses of EBs provide model-independent masses that can be used to more accurately constrain the parameter space of pulsations in the HR diagram. Thus the impact of binary interaction and the excitation physics of pulsations among early-type stars is now possible to study thanks to high-precision TESS data for a large sample of massive stars.


\section*{Data availability}

All data underlying this article are available in the MAST archive ({https://mast.stsci.edu/portal/Mashup/Clients/Mast/Portal.html}).

\section*{Acknowledgements}

We thank Pierre Maxted for alerting us to HD 217919.
The TESS data presented in this paper were obtained from the Mikulski Archive for Space Telescopes (MAST) at the Space Telescope Science Institute (STScI).
STScI is operated by the Association of Universities for Research in Astronomy, Inc., under NASA contract NAS5-26555.
Support to MAST for these data is provided by the NASA Office of Space Science via grant NAG5-7584 and by other grants and contracts.
Funding for the TESS mission is provided by the NASA Explorer Program.
This research has made use of the SIMBAD database, operated at CDS, Strasbourg, France; the SAO/NASA Astrophysics Data System; and the VizieR catalogue access tool, CDS, Strasbourg, France.
DMB gratefully acknowledges funding from the Research Foundation Flanders (FWO) by means of a senior postdoctoral fellowship with grant agreement no.\ 1286521N.


\bibliographystyle{mnras}
\bsp \label{lastpage} \end{document}